\documentclass[
 reprint,
superscriptaddress,
 amsmath,amssymb,
 aps,
prstab,
]{revtex4-1}

\usepackage{graphicx}
\usepackage{dcolumn}
\usepackage{bm}
\usepackage[mathlines]{lineno}

\begin{document}

\preprint{APS/123-QED}

\title{Asymmetric Dual Axis Energy Recovery Linac for Ultra-High Flux sources of coherent X-ray/THz radiation: Investigations Towards its Ultimate Performance}

\author{R. Ainsworth}
 \affiliation{John Adams Institute at University of Oxford, Oxford, UK}
 \author{G. Burt}%
\affiliation{Cockcroft Institute, Lancaster University, Lancaster, UK}
 \author{I. V. Konoplev}
 \affiliation{John Adams Institute at University of Oxford, Oxford, UK}
\author{A. Seryi}
 \affiliation{John Adams Institute at University of Oxford, Oxford, UK}

\date{\today}

\begin{abstract}
In order for sources of coherent high brightness and intensity THz and X-Ray radiation to be accepted by university or industrial R$\&$D laboratories, truly compact, high current and efficient particle accelerators are required. The demand for compactness and efficiency can be satisfied by superconducting RF energy recovery linear accelerators (SRF ERL) allowing effectively minimising the footprint and maximising the efficiency of the system. However such set-ups are affected by regenerative beam-break up (BBU) instabilities which limit the beam current and may terminate the beam transport as well as energy recuperation. In this paper we suggest and discuss a SRF ERL with asymmetric configuration of resonantly coupled accelerating and decelerating cavities. In this type of SRF ERL an electron bunch is passing through accelerating and decelerating cavities once and, as we show in this case, the regenerative BBU instability can be minimised allowing high currents to be achieved. We study the BBU start current in such an asymmetric ERL via analytical and numerical models and discuss the properties of such a system.    
\end{abstract}

\pacs{Valid PACS appear here}
\maketitle


\section{\label{sec:level1}Introduction\protect\\}

The next generation light sources are to be compact, highly efficient, have high repetition rates and high-brilliance radiation pulses. One of the candidates to satisfy all these requirements are light sources based on energy recovery linac (ERLs) \cite{Merminga:2003sr} driven by photo-injection sources. Linacs parameters such as: emmitance, repetition rate and bunch charge, in this case are driven by an electron photo-injector  and laser technologies. Both these technologies have improved dramatically in the last twenty years and such linac drivers capable of generating femtosecond pulses can be routinely bought from specialised companies.  To generate a high-power, high brilliance beam either in THz or X-ray ranges, a high charge electron beam is required and new developments are now bringing Ampere class injectors to reality \cite{PhysRevSTAB.16.083401}. The increase of the bunch charge will lead to an increase of photon yield and brilliance during either x-ray Compton scattering \cite{PhysRevSTAB.8.100702} or generation of THz radiation \cite{PhysRevA.84.013826}. The power ranges of a RF power supply required to drive 100mA (a typical current in such accelerators) are such that energy recovery is required to meet the demand for energy efficient systems. However, adding an energy recovery stage, while increasing the beam charge and repetition rate, leads to the appearance of so called beam break-up (BBU) instabilities \cite{PhysRevSTAB.7.054401}. These instabilities result in beam trajectory shifts, energy recovery degradation, and ultimately, termination of the beam transportation. The regenerative BBU instability is especially damaging to ERL systems and originates from parasitic excitation of transverse higher order modes (HOMs) inside the cavities. The use of the same cavity \cite{PhysRevSTAB.11.032803} or strongly coupled cavities \cite{Feldman:1987nk,wang:2007} means that the positive feedback between the transverse momentum imparted by the HOM, and hence beam displacement and the HOMs amplitude is readily established. A circulating beam through such a system results in a growth in the beam displacement and dephasing for each bunch.  

\begin{figure*}[htbp] 
   \centering
   \includegraphics[width=1\textwidth]{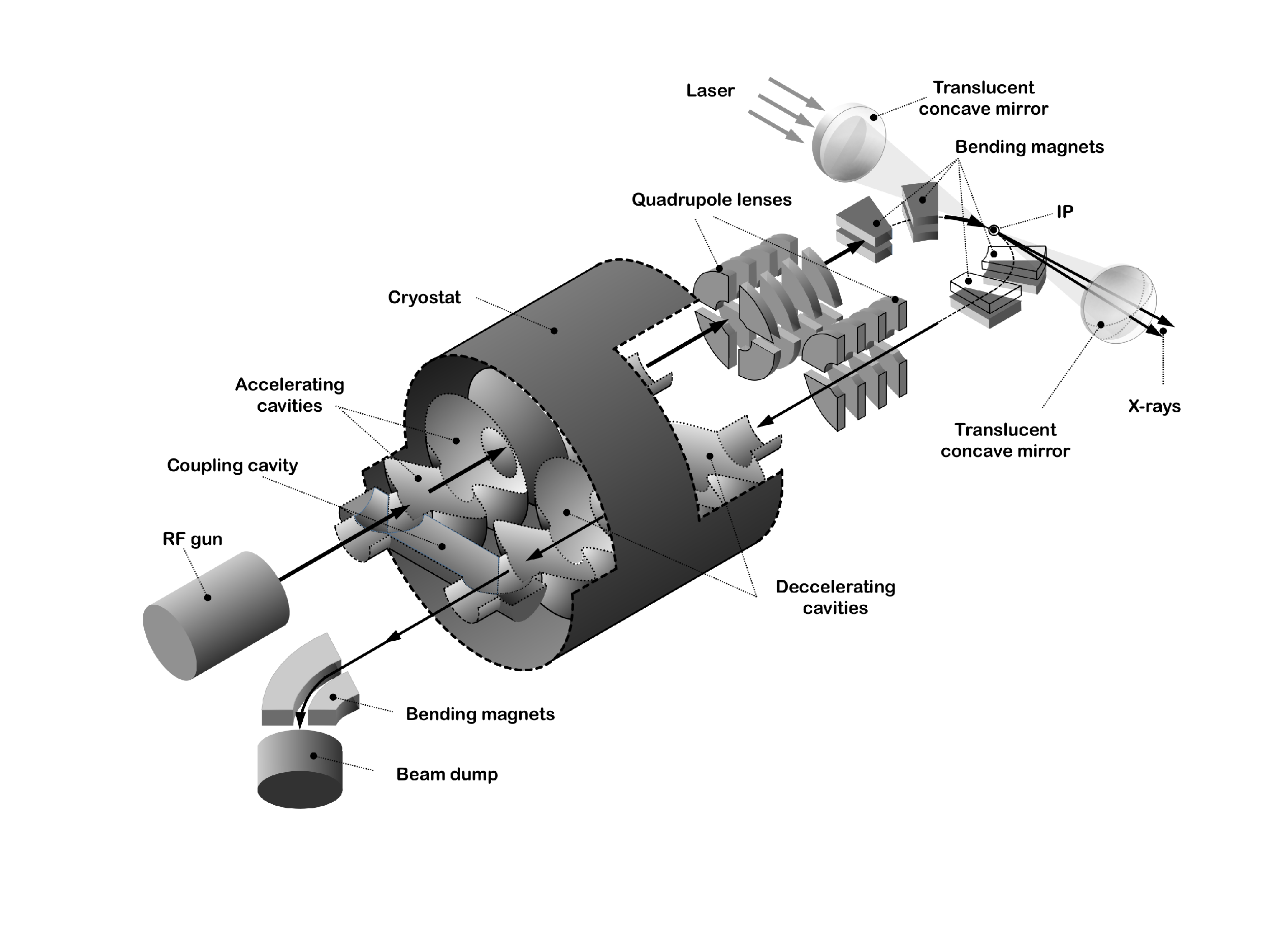} 
   \caption{A schematic of possible single turn ERL system.}
   \label{fig:setup}
\end{figure*}

In this paper we discuss a single turn SRF ERL system \cite{patent1,patent2,patent3} where the beam is transported through the accelerating section, interaction point (IP) and deceleration section only once. This resembles a similar recirculation layout as used in nuclear medicine in the form of the reflexotron developed at Chalk River Nuclear Laboratories \cite{4328851}, but has important differences as we will describe below. In this model, the beam is accelerated inside the acceleration section while in the deceleration section most of the beam energy is extracted and guided through a resonantly coupled section back into the acceleration section. In FIG. \ref{fig:setup}, we show a schematic illustration of a compact source of coherent radiation driven by such a single turn SRF ERL. Both sections consist of the same number of cells but adjusted in such a way that insures that only the operating mode of both sections are fully overlapping each other creating a single operating mode of the cavity, while the HOMs are separated in the frequency domain. 

The two sections are linked by a resonant coupler, and here the resonant coupling means that the two sections are only strongly coupled at the set of frequencies which are overlapping eigen-frequencies of all the three components of the system, i.e. the coupling cell and both sections. There is still some field leakage from one section to another but as it will be shown, the effect is relatively small. Due to the approach described, we reduce the possibility for the multi-pass-regenerative BBU feedback mechanism (where the HOM in the accelerating pass is driven by the field generated by the beam in the decelerating pass) to be established. A potential issue which may be caused by the asymmetry of the structure is that HOMs may have different electrical centres and the beam cannot be at the $E_z$ field nulls for all the HOMs. This theoretically may allow an on-axis bunch to excite a multi-pole HOM which can deflect the following bunch. 

The paper is organised in the following manner: in section \ref{sec:basicDescription} we present the model and basic description of the asymmetric cavity ERL, then in section \ref{subsec:traversedeflect} we discuss the model of bunch trajectories in such ERL and discuss the BBU instabilities. Section \ref{sec:rlc} deals with analysis of HOMs and their start current using the RLC circuit approach. Section \ref{sec:numericalModelling} is dedicated to numerical modelling and to an estimation of BBU start currents for a proposed source. In conclusion (section \ref{sec:conclusions}) we discuss the advantages of the system presented.    

\section{Model and Basic description of asymmetric cavity for ERL}\label{sec:basicDescription}

To increase the efficiency of linac based sources of coherent radiation, electron beam energy recovery is required. Such devices are known as energy recovery linacs (ERLs) and they are assumed to be attractive drivers for compact, energy efficient sources of THz and X-ray radiation. One of the major issues in conventional ERLs is the so called `beam break-up instability' (BBU) which significantly limits the current transported through the system. In ERLs, the dominant instability is the multi-pass regenerative BBU instability where a transverse kick to the beam is given by a higher order mode (HOM). In this paper we suggest a design of a single turn ERL consisting of an accelerating and decelerating section, which are resonantly coupled. 

\begin{figure}[htbp] 
   \centering
   \includegraphics[width=1\columnwidth]{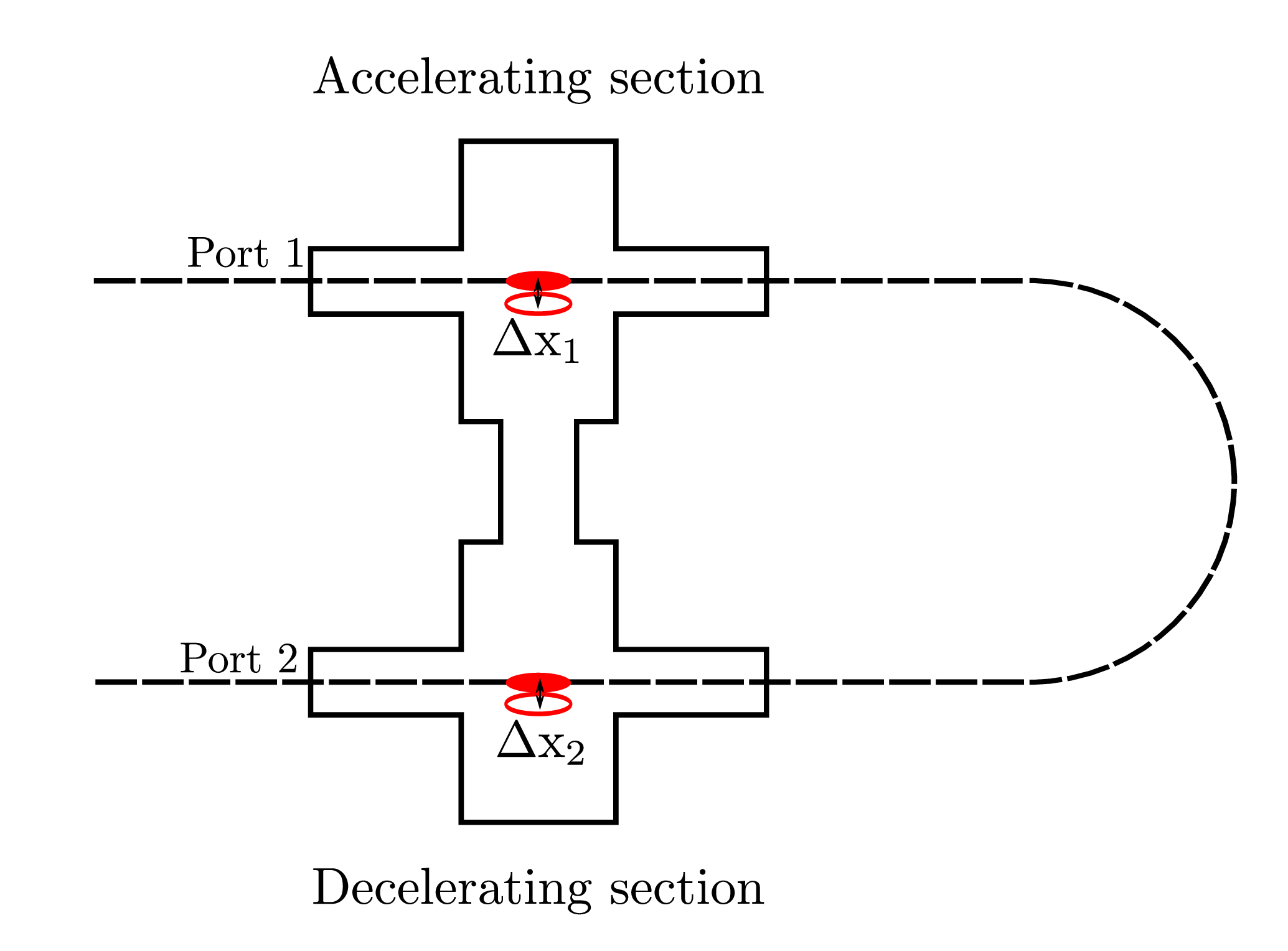} 

   \caption{Schematic of asymmetric ERL structure with accelerated and decelerated bunches. The bunch without red filling indicated the bunch which deviated from the central line trajectory. }
   \label{fig:schematic}
\end{figure}

The ERL under consideration has two axes (FIG \ref{fig:schematic}). An electron bunch will propagate along the first axis and be accelerated. An electron bunch propagating along the second axis will be decelerated and feed energy back into the ERL. The cells' shapes on each axis are tuned to insure that only the operating mode is common for both sections while the higher order mode spectra are different, i.e. the frequencies and Q-factors of the HOMs are different. Due to the resonant coupling between cells located on different axes, the voltage on each axis is nearly the same for the operating mode only, while it varies strongly for all non-overlapping higher order modes. Indeed, the cells act as a single structure at a common resonant frequency (in our case it is the frequency of the operating mode) but they behave as separate structures at most other frequencies. This allows decoupling the accelerating and decelerating structures for HOMs, thus breaking the positive feedback loop. 

Let us first define the model and the approach we use. When a bunch passes through a cavity it is decelerated by the self-induced voltage

\begin{equation}\label{eq:voltage}
V_q = \int_l \vec{E}(\vec{r},z,t) \mathrm{d}\vec{l}
\end{equation}

\noindent where d$\vec{l}$ is defined along the beam trajectory and $\vec{E}(\vec{r},z,t)$ is the vector of the electric field seen by the beam. In the asymmetric system discussed the voltage induced is in general different for each axis, and will be referred as $V_q^{1,2}$ with superscripts indicating the axis number. The cavity geometry parameter $R/Q$ which influences its performance is presented as

\begin{equation}\label{eq:RQintro}
R/Q_{1,2} =  \frac{\left(\int_l \vec{E}(\vec{r})e^{i\omega l/c}\mathrm{d}\vec{l}~\right)_{1,2}^2}{\omega \epsilon \int_V |\vec{E} (\vec{r})|^2 \mathrm{d}\vec{V}}\left(\frac{c}{\omega r}\right)^{2m}
\end{equation}

\noindent where $m$ is the number of full wave azimuthal variations. The case where $m=0$ corresponds to the monopole mode. The parameter $R/Q$ varies from one axis to another as we are considering asymmetric cells. In a conventional symmetric system, which is either made of two identical sections, or the same cavity is used for acceleration and deceleration, the $R/Q$ parameter is constant. Considering that the instantaneous cavity energy change is equal to the instantaneous bunch energy change, one can write (for the case of a single bunch inside the cavity):    

\begin{equation}
\Delta U_q = -\Delta U_c = \left(\frac{c}{\omega r}\right)^2 \frac{\left|V_q^j\right|^2}{2\omega}\frac{1}{R/Q_j}
\end{equation}

\noindent where $j=1,2$ indicates the number of the axis. For the particular case of the structure suggested, the cavity energy gain may vary if the modes' eigen-fields at the axis 1 and 2 are different. The bunches moving along $z$ will interact effectively with the modes having electric field components collinear with the beam trajectory. The dipole modes with the transverse electric and magnetic field components can contribute to the transverse momentum especially if there is transverse offset of the beam trajectory. The transverse momentum can also be gained via interacting with magnetic field of the HOM generated for instance by the previous bunch. If the bunch trajectory deviates from the designed trajectory which we will refer as $r=\sqrt{x^2 +y^2}=0$) the energy lost by the bunch to the HOMs will increase and the bunch will move further away from the axis. In this paper we are considering a small assymmetry of the cells, i.e. the cells on different axes have different shapes, however the variations of shapes are small and leading only to the shift of eigenmodes frequency position while the uncoupled modes' transverse field structures will be considered to be the same. This also means that the same HOMs will be excited by the bunch inside of the accelerating and decelerating structures  but at different frequencies and thus no positive feedback loop will be possible via the coupler (modes are fully separated and not overlapping). As the two structures have no feedback loop between them, no beam instability (in conventional terms) will develop. However, those fields can still provide a large enough transverse kick to cause the bunch to hit the beam-pipe walls.  In addition there is also a single-pass regenerative BBU where an instability develops in a single cavity. This takes place when a kick at the entrance of the structure causes an offset at the exit of the structure, thus exciting a large dipole wake meaning that the following bunch will experience this wake field and its trajectory may deviate from the designed trajectory even further, leading to degradation of the energy recovery or even break-up of the bunch transportation.   

In FIG. \ref{fig:schematic}, the schematic of possible bunch trajectories are shown with $r=r_0 =0$ representing the designed trajectory while $\Delta x$ deviation from the designed trajectory. A `transparent' bunch in this figure illustrates a deviation of possible trajectory from the `design orbit' (corresponds to a bunch motion along both central lines) which can be due to the excitation of high order modes or trajectory perturbation at the interaction point (IP). Further in the text, considering azimuthal symmetry of the system and to simplify notations (without losing generality) we consider only the $xz$ plane (i.e. $y=0$ and $r=x$) resulting in $r(t+t_0)=x(t+t_0)= \mathbf{R_{11}}x_0 + \mathbf{R_{12}} \theta$.

\subsection{Transverse deflection of a single bunch} \label{subsec:traversedeflect}

The deviation from the `design orbit' can be calculated using

\begin{equation}\label{eq:rtplust0}
x(t+t_0) = \mathbf{R_{11}}x_0 + \mathbf{R_{12}} \theta
\end{equation}

\noindent where $t_0$ is the time required for an electron to go from port 1 to port 2, $\theta$ is the angle gained by electrons at the cavity exit, $\mathbf{R_{11}}$ and $\mathbf{R_{12}}$ are the elements of the transport matrix that relates the angle at the cavity exit and bunch transverse position $x$ at port 2.

For simplicity but without losing generality we assume that at the entry point all bunches are having finite but negligible (in the first approximation) dimensions and move along the central axis. The finite transverse dimension of the bunch will result in generation of transverse momentum leading to the shift of some electrons from the `design orbit' of the system defined by the transport $\mathbf{R}$ matrix (we are ignoring for now some high order corrections associated with $\mathbf{R_{11}}$ term) 

\begin{equation}\label{eq:rtplust0sim}
x(t+t_0) = \mathbf{R_{12}} \theta
\end{equation}

 Here, for clarity only, we neglect the coupling of the offset and vertical/longitudinal beam parameters and assume that d$\vec{l}=$d$\vec{z}$. Taking into account that the beam is relativistic and its longitudinal velocity is close to speed of light, we can estimate the angle at which the bunch leaves the cavity as:

\begin{equation}
\theta \cong \frac{eV_{\perp}}{eW}\cong\frac{V_{\perp}}{W}
\end{equation}

\noindent where ($eW$) is the full bunch energy and $V_\perp$ is the effective transverse voltage seen by the bunch. Using (\ref{eq:rtplust0sim}) and taking into account that $x_0 =0$ one gets the expression for the radial deviation of the bunch from the second axis 

\begin{equation}\label{eq:deltar_0}
\Delta x = \mathbf{R_{12}} \theta \cong \mathbf{R_{12}} \frac{V_\perp}{W}
\end{equation}

\noindent We note that if we use $(y,z)$ plane it would transfer into $\Delta y = \mathbf{R_{34}} \theta \cong \mathbf{R_{34}} \frac{V_\perp}{W}$

The effective transverse potential which leads to the change of the total transverse momentum and thus trajectory deviation can be expressed as  

\begin{equation}\label{eq:VperpFull}
V_\perp = -c\int^T_0 \mathrm{d}t \int^L_0  \nabla_\perp E_z (r,\varphi, z, t) \mathrm{d} z
\end{equation}

\noindent where $\nabla_\perp = \frac{\partial}{\partial r}\vec{r} + \frac{1}{r}\frac{\partial}{\partial \varphi}\vec{\varphi}$.

Here for convience we keep cylindrical coordinates however if inserted into (\ref{eq:deltar_0}) a coordinate change may be required.
This introduced potential is observed by the electrons passing through the accelerating gap $L$ over time $T$. To evaluate $V_\perp$ and thus the bunch deviation accurately, a full 3D eigenmode analysis is required. Here we will make few approximations to estimate a single bunch trajectory deviation. First of all, we take into account monochromaticity of the field i.e. $\propto e^{-i\omega t}$ then (\ref{eq:VperpFull}) can be rewritten as:	

\begin{equation}
V_\perp = -\frac{2c}{\omega} e^{-i\phi_T}\sin\phi_T \left[ \int^L_0 \nabla_\perp E_z (r, \varphi, z) \mathrm{d} z\right]
\end{equation}

\noindent where $\phi_T=\omega T/2$ is a half transient phase. As we are looking at the maximum kick observed by the electrons it is clear that it can be achieved if $\phi_t=\pi/2$, so the transverse potential becomes

\begin{equation}
V_\perp = -i \frac{2c}{\omega} \left[ \int^L_0 \nabla_\perp E_z (r, \varphi, z) \mathrm{d} z \right]
\end{equation}

\noindent indicating that the effective transverse potential is $\pi/2$ phase shifted from the longitudinal (accelerating) field. In the case of multipole fields, the following convention is used to describe the longitudinal electric field: $E_z(r,\varphi,z) = 1/2 \Re(E_z(r,z))\cos(m\varphi)$. The polarisation axis is chosen such that the maximum electric field is at $\varphi = 0$. At this point a coordinate system change is also made to cartesian coordinates for simplicity and to match (\ref{eq:deltar_0}). Making use of (\ref{eq:voltage}) and assuming $\varphi=0$ such that the bunch see the maximum field excitation, the transverse potential can be written in terms of the longitudinal voltage

\begin{equation}
V_\perp = -i \frac{c}{\omega} \frac{\partial V_\parallel}{\partial x}
\end{equation}

We also assumed that a bunch propagating along the `design orbit' is perfectly timed to `see' both the  maximum accelerating and decelerating potentials. Any trajectory perturbation due to interaction at IP is ignored. Assuming that bunches arrive to port 1 on axis, the trajectory shift at the second port is given by:

\begin{equation}\label{eq:deltar}
\Delta x =  \mathbf{R_{12}} \frac{-i\frac{2c}{\omega}\left[\int^L_0 \frac{\partial}{\partial x} E_z (x, z) \mathrm{d}z\right]}{W}
\end{equation}

The expression (\ref{eq:deltar}) gives an opportunity to make a first estimation of an upper bound for the maximum $\mathbf{R_{12}}$ parameter knowing some basic properties of the cavity and assuming that $E_z (x,z)\propto \cos(k_z z)$ (pill-box like cavity). Evaluating the integral in (\ref{eq:deltar}) we get: 

\begin{equation}\label{eq:deltar_pillbox}
\Delta x \cong -i\frac{2c}{\omega}\left[ \frac{\partial}{\partial x}  V(x)\right]\frac{L}{W}\left(\frac{\sin\phi_z}{\phi_z}\right)\mathbf{R_{12}}
\end{equation}

\noindent where $\phi_z=k_z L$ is the electron bunch phase shift at the cavity exit and $k_z= \sqrt{(\omega/c)^2-k_\perp^2}$. We assume that $k_z = 0$ as in this case $\phi_z = k_zL=0$ and function $\sin x/x$ in (\ref{eq:deltar_pillbox}) has maximum at this value of $x$. Thus, for all other values of $k_z$, the function will be smaller and $\Delta x$  will takes values from 0 to $\Delta x$ as defined in (\ref{eq:deviation})

\begin{equation}\label{eq:deviation}
\Delta x \cong -\frac{2c}{\omega}{ \frac{\partial}{\partial x}  V(x)}\frac{L}{W} \mathbf{R_{12}}
\end{equation}

This expression shows the link between bunch's trajectory deviation and machine's parameters. Therefore, to reduce the deviation one may either  reduce the parameter  $\frac{2c}{\omega}\frac{L}{W}$   or limit the $\mathbf{R}_{12}$ parameter for a given aperture (diameter $D_2$) of deceleration cavity (port 2) as shown

\begin{equation}\label{eq:r12cond1}
\mathbf{R_{12}} \ll D_2 W\frac{k}{2k_zL^2}\frac{1}{| \frac{\partial}{\partial x}  V|}
\end{equation}

One can further evaluate the expression by assuming  polynomial representation \cite{Navarro-Tapia:2013bpa} of the potential, i.e.  $V=\sum\limits^M_{m=0} x^m V_{m+1}$  where $m=0$ is associated with the monopole mode and $m=1$ with the dipole mode (we ignore the other modes based on the assumption of small deviation from the `design trajectory'). Let us recall that we assumed $r=x$. Substituting this into (\ref{eq:r12cond1}) one gets  

\begin{equation}\label{eq:r12cond2}
\mathbf{R_{12}} \ll D_2 W\frac{k}{2k_zL^2}\frac{1}{m\left|\sum\limits^M_{m=1} x^{m-1} V_{m+1}\right|}
\end{equation}

It is clear that this condition is not sufficient for the ERL to operate as the bunch with a trajectory deviated from the `design orbit' will be delayed due to the longer path given by the transport matrix element $\mathbf{R_{52}}$ and its deceleration (energy recovery) will be affected, leading to possible interruption of ERL operation even if the bunch passed through the decelerating section. If the beam enters the decelerating cavity with a deviation of the transverse position $x_0+\Delta x$ then the transient time $T_g$ from acceleration to deceleration sections is changed as ($T_g + \Delta \tau$) and deceleration will shift from its optimum. This phenomenon is a bunch de-phasing and knowing $\Delta x$  one may estimate the bunch time deviation $\Delta \tau$ from the time travel along the design trajectory $T_g$ due to trajectory deviation $\Delta x$. Ignoring transport matrix elements such as $\mathbf{R}_{52}$ but keeping $\mathbf{R}_{12}$, i.e. neglecting any betatron oscillations in the transport line and assuming that $\delta x \gg S_0$, where $S_0$ is the length of the design trajectory, and that the electron bunch travels with the speed of light $c$, one finds that the total length of the deviated trajectory is 

\begin{equation}
S = \sqrt{S_0^2 + (\mathbf{R}_{12}\theta)^2} \cong S_0 \left[ 1+ \frac{1}{2} \left(\frac{\mathbf{R_{12}}\theta}{S_0}\right)^2\right]
\end{equation}

\noindent while the time deviation from an optimum time along the `design trajectory' is 

\begin{equation}\label{eq:deltatau}
\Delta \tau \cong \frac{1}{2} \left(\frac{\mathbf{R_{12}}\theta}{S_0}\right)^2 T_g = \frac{\Delta x}{2S_0}\frac{\Delta x}{c}
\end{equation}

Multiplying (\ref{eq:deltatau}) by the frequency of the operating mode $\omega_0$ and taking into account that $\omega_0 T_g=2\pi n$ where $n$ is the number of field oscillations in the cavity during the bunch travel from the accelerating to the decelerating arm, (\ref{eq:deltatau}) can be transformed:

\begin{equation}\label{eq:deltaphi1}
\Delta \phi \cong \left(\frac{\mathbf{R_{12}}\theta}{S_0}\right)^2 \pi n
\end{equation}

It is clear that if there is no deviation there is no phase variation from the optimal value. Noting that \textit{n} has the meaning of the number of wavelength of operating modes $\lambda_0=2\pi c/\omega_0$ along the optimal path $S_0$, we can present (\ref{eq:deltaphi1}) in a slightly different form 

\begin{equation}\label{eq:deltaphi2}
\Delta \phi \cong \frac{(\mathbf{R_{12}}\theta)\Delta x}{S_0\lambda_0} \pi
\end{equation}

Taking into account that $\Delta\phi\ll\pi$, one can write a second condition for $\mathbf{R_{12}}$ in order to guarantee a small deviation of $\Delta\phi$ and thus effective deceleration. We also note that to avoid beam dumping before deceleration cavity the condition $\Delta x < D$ needs to be satisfied and thus we can rewrite (\ref{eq:deltaphi2}) as 

\begin{equation}
\Delta \phi \cong \frac{(\mathbf{R_{12}}\theta)\Delta x}{S_0\lambda_0} < \frac{D^2}{S_0\lambda_0}\pi \ll \pi
\end{equation}

In conventional accelerators condition (\ref{eq:r12cond2}) is ``stronger'' and thus sufficient. The effective transverse potential due to multiple bunches can also be derived. The derivation can follow the approach presented in \cite{Burt:2007zzd} and can be found in the Appendix \ref{app:multilaunch}.

\section{HOMs overlap and BBU start current }\label{sec:rlc}

\begin{figure}[htbp] 
   \centering
   \includegraphics[width=1\columnwidth]{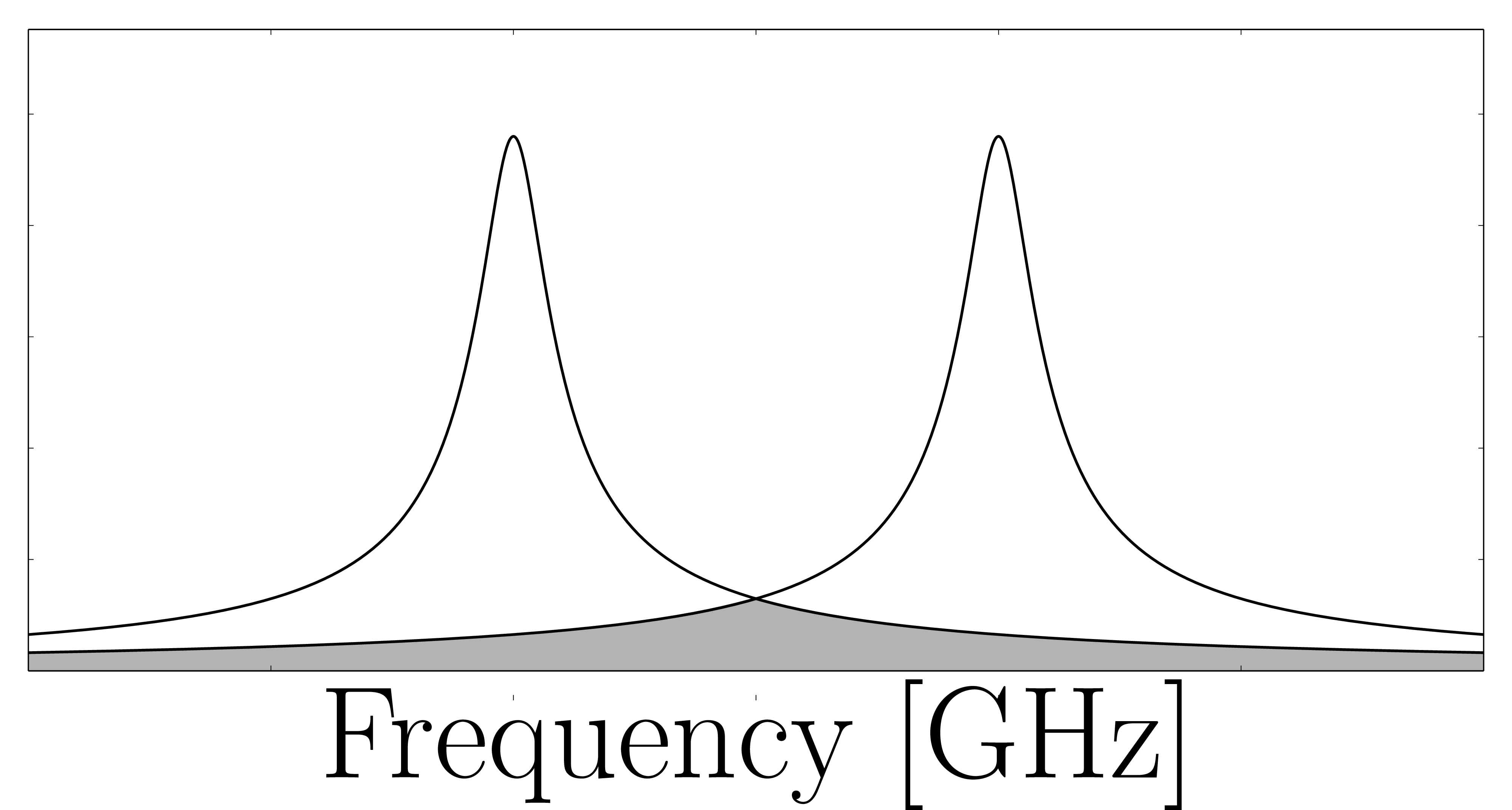} 
   \caption{An example of two modes with different frequencies which overlap due to their finite bandwidth. The shaded region is helping illustrate where two modes overlap but does not indicate the magnitude of coupling.}
   \label{fig:overlap}
\end{figure}

    In this section we will employ RLC circuit approach to describe the system. Let us now consider that in reality the HOMs of the accelerating and decelerating sections will overlap due to their finite bandwidths. In FIG. \ref{fig:overlap} a schematic of such an overlap is shown. At the shaded (overlap section) part of the spectra one may expect a leakage of the energy from one arm of the cavity into another. This will lead to a coupling of energy between the accelerating and decelerating sections, providing a feedback path for instabilities to develop. One such instability is multi-pass regenerative BBU. 

Here we meet our first issue when trying to describe a dual axis machine. The standard RLC equations are not valid as the impedance is dependent on which cavity axis the beam traverses. However, for a given energy and for a given mode the ratio between the voltages on each axis is constant, a scaling factor can be introduced. In essence the system acts like a transformer with a load and a current source at both the primary and secondary axis, with a ratio of $N_{SP}$. Considering the voltage at the accelerating section $V_1$ and at the decelerating section $V_2$ for a given stored energy we get

\begin{equation}
V_2 = N_{SP} V_1
\end{equation}

\begin{figure}[htbp] 
   \centering
   \includegraphics[width=1\columnwidth]{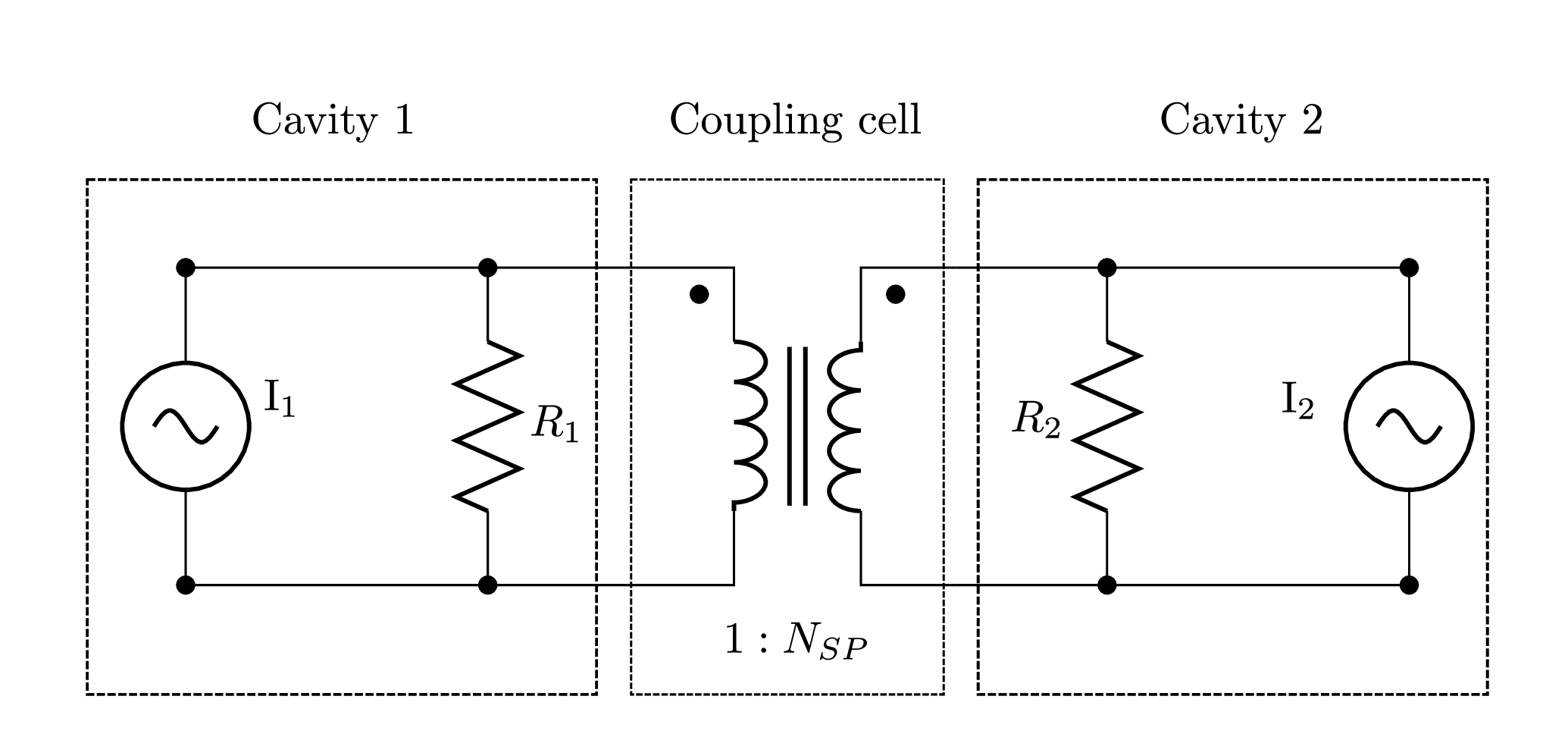} 
   \caption{A circuit diagram of the dual axis structure.}
   \label{fig:cavityCircuit}
\end{figure}

From the primary current source the voltage of the secondary is stepped either up or down depending on the transformer ratio. In this context, it is useful to define the voltage in the decelerating section as seen from the accelerating section such that we can view the system as two parallel circuits.  The voltage on the second axis, $V_2$, when looked at from the primary axis (i.e. the bunch propagates along first axis) is transformed to $V_2^\prime$ which will be the same as $V_1$ and therefore

\begin{equation}\label{eq:v1}
V_1 = V_2^\prime = \frac{V_2}{N_{SP}}
\end{equation}

First lets calculate the shunt impedance of each structure, ignoring the losses in the other structure. Taking into account that the shunt impedance of the uncoupled acceleration and deceleration sections are the same, $R_1 \cong R_2$, this leads to the shunt impedance of each structure, if coupled, to be  different when seen from the other axis. Indeed, the impedance of the secondary structure (located on the 2nd axis) when viewed from the source on the first axis is $R_2^\prime$, and can be written as \cite{hughes1954fundamentals}

\begin{equation}
R_2^\prime =R_1 N_{SP}^2 = R_2 /N_{SP}^2
\end{equation}

Also the beam across one accelerating/decelerating gap can be looked at as being a current source in parallel with both impedances $R_{1,2}$ (loads). In this case the beam will apply voltages across the gaps while the stored energy, and the losses will be also split between the both loads. Such a description of the system allows one to define the total impedance of the whole system as a sum of two parallel impedances  and find the voltage on the first axis, $V_1$, and second axis, $V_2$, induced due to the bunch propagation along  the first/second axes. The expression for $V_1$ can be calculated from (\ref{eq:v1}) and is given in this case as: 

\begin{equation}
V_1 =   I_1 \frac{R_1}{1+N_{SP}^2}
\end{equation}

\noindent And the voltage across the decelerating gap due to the beam in the accelerating gap is similarly given by

\begin{equation}
V_2 = N_{SP}V_2^\prime = N_{SP}V_1 = \frac{I_1R_1N_{SP}}{1 + N_{SP}^2}
\end{equation}

The total effective transverse impedances for both gaps (combined in parallel) $R_{\perp 1}$ and $R_{\perp 2}$ can hence be introduced for the accelerating and decelerating gaps respectively

\begin{equation}
R_{\perp 1} = \frac{R_1}{1+N_{SP}^2}
\end{equation}

\begin{equation}
R_{\perp 2} = N_{SP}R_{\perp 1}=\frac{N_{SP}R_1}{1+N_{SP}^2}
\end{equation}

Let us apply these expressions to the calculation of multi-pass regenerative BBU. A bunch traversing the accelerating gap of the cavity will experience a transverse force due to the previous bunches excitation of a dipole HOM and will deposit energy into the cavity. It will  experience an acceleration or deceleration dependent on the offset of an individual bunch from the electrical centre of the cavity, $x_a$. The sum wakefield due to the previous bunches (we are looking at a single dipole HOM) has amplitude $V_d$, phase $\phi$ and the accumulated energy change of the bunch in terms of voltage is $\Delta U_1$,

\begin{equation}
\Delta U_1 = q x_a V_d \cos\phi 
\end{equation}

Let us note that the phase $\phi$ is between the bunch centroid and the peak accelerating EM field excited by the previous bunches The trailing electron bunch will have a deflection which is defined by an effective transverse voltage (as discussed above) and is given in this case as: 

\begin{equation}
V_{\perp 1} = \frac{c}{\omega} V_{d} \sin \phi
\end{equation}

We note that the transverse effective potential and decelerating voltage are $-\pi/2$ shifted in respect to each other (term -i in front of the expression) as given by Panofsky-Wenzel theorem \cite{p-w}. A bunch deflected by this effective potential will arrive at the decelerating gap with a transverse offset from the design orbit, $\Delta x$, which (as we discussed) can be calculated for a given beam energy $W$, using the transfer matrix $\mathbf{R}$,  

\begin{equation}
\Delta x \cong \mathbf{R_{12}}\frac{V_\perp}{W}\sin \phi
\end{equation}

Similarly as we done above, we assumed that $\mathbf{R}_{11}x_a$ is small and therefore, neglect this term. Since the two gaps are coupled, the dipole mode also has a longitudinal voltage component in the decelerating gap leading to acceleration or deceleration of the bunch, and the maximum energy $\Delta U_2$, deposited in the mode by the beam traversing the decelerating gap, is given by

\begin{equation}
\Delta U_2 = q(x_d + \Delta x) N_{SP}V_{d}C(T_g)
\end{equation}

\noindent where $C(T_g)=\cos (\phi + \omega T_g)$, $x_d$ is the distance between the HOM electrical centre and the design orbit in the decelerating axis (in general $x_a\neq x_d$), $T_g$ is the time travel from the accelerating to the decelerating cavity. We assume that the decay of the field in the cavity between the bunches going through each axis is small. For a pure dipole mode, if $V_d$ grows in time then so does the bunch offset $\Delta x$, and hence to calculate the BBU threshold current we only need to work out the power balance in order to derive the condition for the instabilities not to grow. The average power over one RF cycle (assuming every bucket is filled) is

\begin{equation}
I_0V_{d}x_a\cos\phi + I_0(x_d + \Delta x)N_{SP}V_{d}C(T_g) - P_c \le 0 \\
\end{equation}

\noindent where $I_0$ is the beam current, $P_c$ is the energy losses due to the finite Q-factor of the cavity. The first term on the left hand-side of the expression indicates the EM field energy gain inside the accelerating section, while the second term shows the energy change of the EM field inside the deceleration section. It should be noted that for the case of every bucket being filled the 1st and 2nd terms may not necessarily be due to the same bunch. Inserting the bunch offset at the second axis due to the kick on the first axis we obtain

\begin{gather}
\begin{aligned}\label{eq:equalityI0}
&I_0V_{d}x_0\cos\phi + \\
&I_0N_{SP}V_{d}\left[x_d + \mathbf{R_{12}}\frac{V_{d}}{W}\frac{c}{\omega}\sin\phi\right]C(T_g) - P_c > 0
\end{aligned}
\end{gather}

The cavity power losses for a given load are defined by $Q_L$ (Q-factor which takes into account both internal and external losses due to HOM couplers) which is given by

\begin{equation}
P_c = \frac{V_{\perp}^2}{R/Q_{\perp 1}Q_L}
\end{equation}

\noindent where $R/Q_{\perp 1}$ is the transverse $R/Q$ for a dipole mode (as defined above). Inserting this into equation (\ref{eq:equalityI0}) and substituting for $V_\perp$ yields

\begin{gather}
\begin{aligned}
&I_0V_d x_a\cos\phi \\ 
&+ I_0N_{SP}V_{d}\left[x_d + \mathbf{R_{12}}\frac{V_d}{W}\left(\frac{c}{\omega}\right)\sin \phi \right]C(T_g) \\
&- \frac{V_{\perp }^2}{R/Q_{\perp 1}Q_L} > 0
\end{aligned}
\end{gather}

Rearranging for $I_0$ yields

\begin{equation}
I_0 > \frac{cV_\perp}{\omega R/Q_{\perp 1}Q_L\left(x_a\cos\phi+ N_{SP} \chi C(T_g)\right)}
\end{equation}

\noindent where 

\begin{equation}
\chi = \frac{x_d}{V_{\perp 1}} + \frac{\mathbf{R_{12}}}{W}\sin \phi 
\end{equation}

We note that, since $V_\perp$ and $\phi$ are not known, in order to find the solution, a full wakefield calculation and analysis are required in such an asymmetric structure. However, if we consider (just for our case) that both of the offsets on the first axis (i.e. bunch centroid offset and the offset between the electrical centre and the bunch design orbit) are small as compared to the bunch offset at the second axis, then due to the transverse kick near the BBU limit the non-equilibrium current can be simplified to 

\begin{equation}
I_0 > \frac{cW}{N_{SP}\omega \mathbf{R_{12}}R/Q_{\perp 1}Q_LC(T_g)}
\end{equation}

This simplification does however mean that the start current calculated is an over estimate by a factor of approximately $(x_d+a)/a$ where $a$ is the beam aperture. This equation is however still a function of the initial transition phases $\phi$ and $\omega T_g$. Let us look at the worst case (i.e. the largest transverse kick at the first axis when  $\phi=\pi/2$ ), which leads to the following estimate 

\begin{equation}\label{eq:currentWorst}
I_0 > \frac{cW}{N_{SP}\omega \mathbf{R_{12}}R/Q_{\perp 1}Q_L\cos(\omega T_g)}
\end{equation}

\noindent If $\omega T_g \cong 2\pi m$ and $\phi=\pi/4$ the equation (\ref{eq:currentWorst}) becomes

\begin{equation}
I_0 > \frac{2cW}{N_{SP}\omega R/Q_{\perp 1}Q_L\mathbf{R_{12}}}
\end{equation}

Expanding this expression using the transformer ratio and the relation between the impedances of a single separate structure (as discussed above and not the effective transverse impedance of the whole cavity) one will get

\begin{equation}
I_0 > \frac{\left(1+N_{SP}^2\right)cW}{N_{SP}\omega \mathbf{R_{12}}R/Q_{\perp 1}Q_L}
\end{equation}

Comparing it with the start current of symmetric cavity, $N_{SP}=1$, one notes that the start current of an asymmetric cavity is increased. In a symmetric cavity each HOM has equal field in each structure ($N_{SP}=1$) leading to the following relation between the currents: 

\begin{equation}
I_{asymmetric} > \frac{\left(1+N_{SP}^2\right)I_{symmetric}}{2N_{SP}}
\end{equation}

Analysing the expression, one notes, that the ratio between the currents has the minimum value of 1 if $N_{SP}=1$  (a symmetric cavity). This can be interpreted in the following way. If $N_{SP} < 1$, the energy feedback from the second cavity is smaller, leading to start current increase (as the denominator is proportional to $N_{SP}$). On the other hand, if $N_{SP}$ is large, the kick on the first axis becomes small and to get BBU, the start current should be increased, which manifests itself in the $N_{SP}^2$ term. One also notes that for $N_{SP}=1$ the cavity has twice the power losses of a standard cavity and thus the start current is doubled compared to a recirculating machine. 

\section{Numerical model of wake field generation in asymmetric cavity }\label{sec:numericalModelling}
A bunch propagating inside the structure excites wake fields which can be described as superposition of eigenmodes of the cavity i.e. $\vec{E}(r,z) = \sum_s C_s E_s$ where $C_s$ represents a coefficient and $E_s$ corresponds to the electric field for the eigenmode with indice $s$. The excitation of a single eigenmode is discussed above and calculation of complex wake field considering all modes is outside the scope of analytical theory and can be done using numerical modeling. The goal of the numerical modeling in this case is to identify the most dangerous modes in the system described as close to the real case as possible, using eigenmode solvers, and using analytics developed to suggest the methods to suppress these modes. The simulations were performed using the ACE3P electromagnetic suite developed at SLAC. Taking into account that such systems were not modelled previously we conducted investigations using both frequency (Omega3P \cite{Lee:2009zzc}) and time (T3P \cite{1742-6596-180-1-012004}) domain approaches. This was done to verify the results observed and illustrate some of the conclusions. For all cases, curvilinear tetrahedral mesh-elements were used.
We have also investigated the wakefield generation. The wakefield was Fourier analysed  and compared with location of eigenmodes of the `cold' system. Taking into account the repetition rates of the system and considering analytics developed, we show the possibility to increase further the BBU start current via the introduction of the additional losses for HOMs.

\subsection{Cavity Design}

\begin{figure}[htbp] 
   \centering
   \includegraphics[width=1\columnwidth]{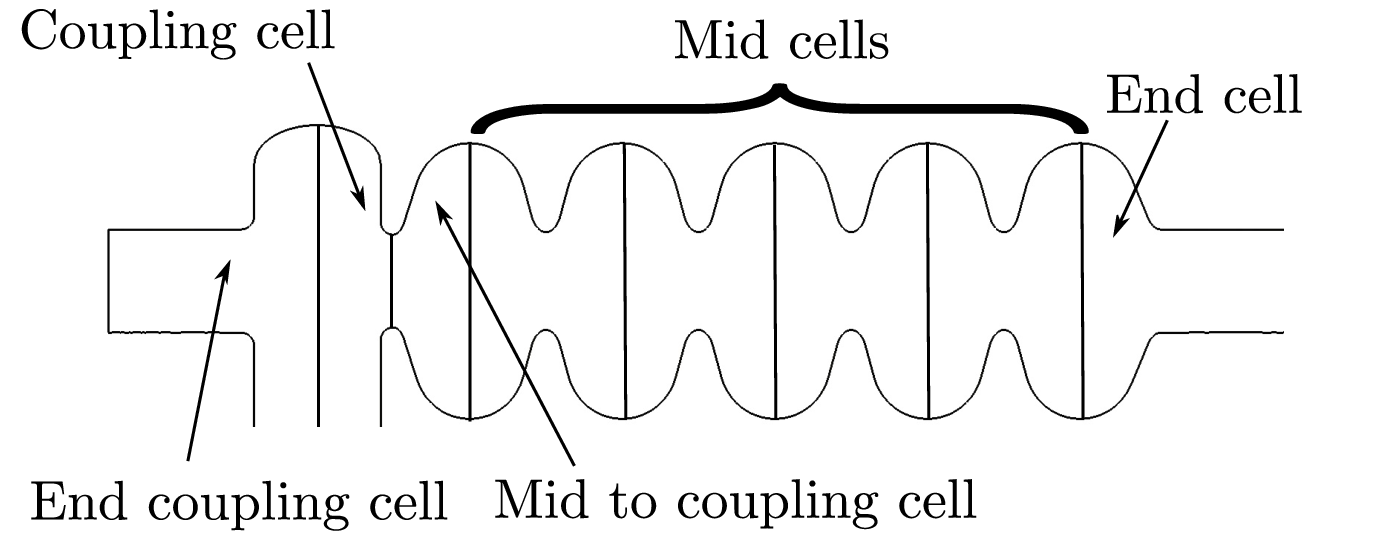} 
   \caption{Schematic of one axis of the structure.}
   \label{fig:axis1}
\end{figure}

Let us consider the design of the ERL structure proposed. The structure is created from two typical elliptical SRF cavities with different parameters joined by a coupling cell. A typical elliptical SRF cavity is formed from a number N of mid-cells plus end-cells to take into account the coupling to the beam pipe. For the asymmetric structure, the design is more complex as shown in FIG. \ref{fig:axis1}. A key step in designing the structure is to make the mid-cells for each axis different but still maintaining the same operating mode.

The design process involved using the mid cell design from a conventional cavity (in this case, the TESLA cavity \cite{PhysRevSTAB.3.092001} shape, operating frequency of 1.3 GHz) and to vary the parameters, finding two shapes which had the biggest difference in geometry yet still shared the same operating mode. The mid-cells are required to have the same frequency and $R/Q$ to ensure equal voltages in both modes. The $R/Q$ does not have to be exactly the same but the frequency must be. A change in $R/Q$ will result in different field amplitudes in the two sections. The length of the half-cell is kept to a quarter of a wavelength such that the designs can be scaled to other frequencies. For example, if the desired frequency was 1.95 GHz, each parameter can be multiplied by a factor of 1.3/1.95. Additional tuning is needed for the end-cells where the length is no longer half a wavelength. Of importance is to try to separate the higher order mode spectrum as much as possible. Once two shapes were found, a dispersion diagram was created for each design and an example is shown in FIG. \ref{fig:dispersion}. Each line on the dispersion represents the passband for a particular mode in the case of an infinitely long periodic structure.  A passband shows the frequency range in which a mode can propagate with a particular phase advance between adjacent cells in a structure. When the structure is used with a finite number of cells, a pass band will split into a finite number of modes that lie on the line. We assume that the optimum scenario is that the pass bands will have a frequency separation on the order of one or a few MHz. 

\begin{figure}[htbp] 
   \centering
   
   \includegraphics[width=1\columnwidth]{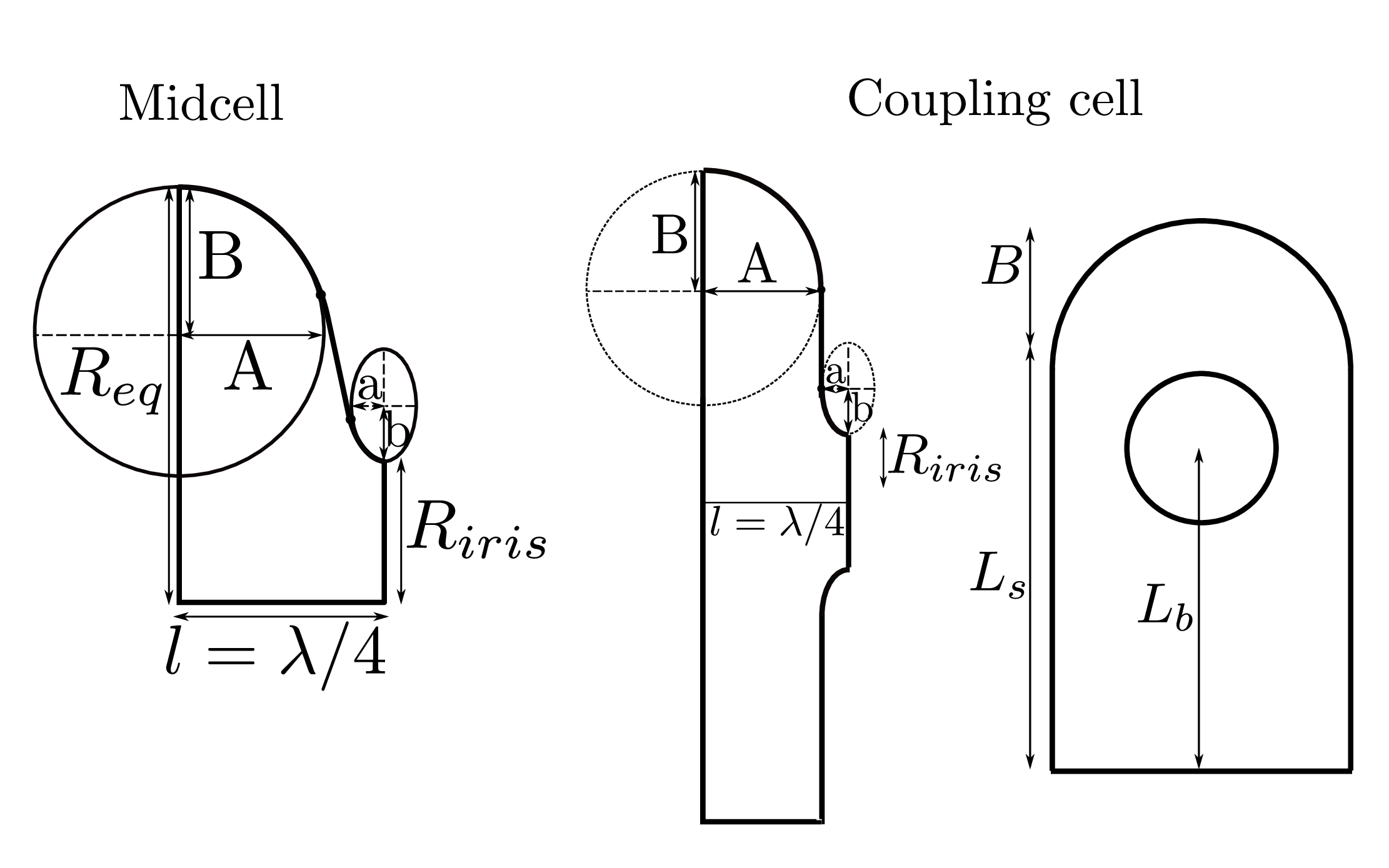}

   \caption{Schematic of the mid cell (left) and the coupling cell (right).}
   \label{fig:cellSchematic}
\end{figure}

\begin{table}[htbp]
   \centering
   \begin{tabular}{lll}
   \hline
   \hline
   Parameter  & Axis 1 cell~[mm] & Axis 2 cell~[mm] \\
   \hline
   \multicolumn{3}{c}{Mid cells} \\
   \hline
   $R_{eq} $& 103.3 & 103.3 \\
   $A$ & 42 & 42 \\
   $B$ & 42 & 43.1 \\
   $R_{iris}$ & 35.75 & 37 \\
   $a$ & 12.75 & 11.75\\
   $b$ & 18 & 20 \\
   $l$ &  57.7 & 57.7 \\
\hline
   \multicolumn{3}{c}{End cells} \\
\hline
 $R_{eq} $& 103.3 & 104.3 \\
   $A$ & 42 & 42 \\
   $B$ & 42 & 43 \\
   $R_{iris}$ & 39 & 39 \\
   $a$ & 12.75 & 11.75\\
   $b$ & 18 & 20 \\
   $l$ &  58.54 & 60.96 \\
   \hline
   \multicolumn{3}{c}{Mid to coupling cells} \\
\hline
    $R_{eq} $& 103.3 & 104.3 \\
   $A$ & 42 & 42 \\
   $B$ & 43.4 & 43.5 \\
   $R_{iris}$ & 35 & 35 \\
   $a$ & 12.75 & 9.69\\
   $b$ & 18 & 20 \\
   $l$ &  57.7 & 57.7 \\
      \hline
   \multicolumn{3}{c}{Coupling cells} \\
\hline
   $A$ & 48.052 & 48.052\\
   $B$ & 29 & 29\\
   $R_{iris}$ & 35 & 35 \\
   $a$ & 9.6 & 9.6\\
   $b$ & 10.152 & 10.152 \\
   $l$ & 57.652 & 57.652 \\
   $L_s$ & 150 & 150 \\
   $L_b$ & 111 & 111\\
  \hline
   \multicolumn{3}{c}{End coupling cells} \\
\hline
   $A$ & 47.5& 47.5\\
   $B$ & 29.76 & 29.76 \\
   $R_{iris}$ & 39 & 39 \\
   $a$ & 9.945 & 9.945\\
   $b$ & 9.945 & 9.945 \\
   $l$ & 57.652 & 57.652 \\
   $L_s$ & 150 & 150 \\
   $L_b$ & 111 & 111\\
   \hline
  \hline
   \end{tabular}
   \caption{Parameters used to construct numerical model.}
   \label{tab:cellParams}
\end{table}

\begin{figure}[htbp] 
   \centering
   \includegraphics[width=1\columnwidth]{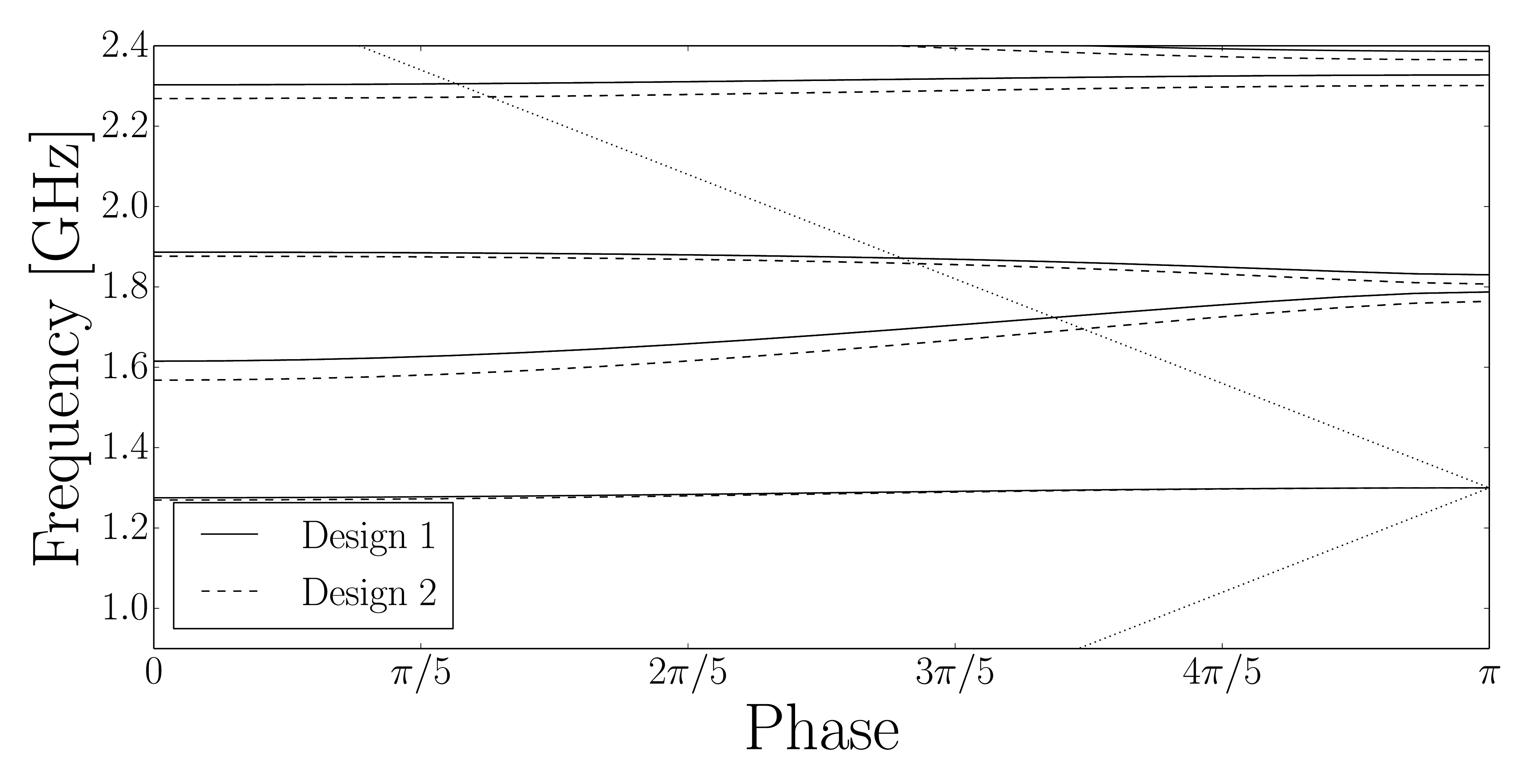} 
   \caption{Dispersion diagram for the two mid cell designs.  A clear separation between the passbands can be seen with the exception of the first passband in which the operating mode is shared. For a symmetric structure the passbands would be identical. The dotted line shows the speed of light curve.}
   \label{fig:dispersion}
\end{figure}

The design of the coupling cell is more complex. Its shape is racetrack-like in order to couple both accelerating and decelerating sections with  the axes of the sections separated by distance $2L_b$ (FIG. \ref{fig:cellSchematic}). The wall angle between the two ellipses is forced to be $90^\circ$. The example is shown in FIG. \ref{fig:cellSchematic} and TABLE \ref{tab:cellParams}. The main purpose of the design is to couple strongly the sections at operating frequency and effectively split them at all the other frequencies at which HOMs are located. 

To create the end cell, the parameters of the mid cell were modified to take into account the beam pipe dimensions. Another change in dimensions is also needed to join the mid cell to the coupling cell. The mid cells for each axis have a different iris radius, however the coupling cell is chosen to be symmetric for simplicity and thus another cell is needed to match the iris radius with the coupling cell. The example of such dimensions is shown in TABLE \ref{tab:cellParams}. One notes that these dimensions can vary depending on the operating mode frequency.

\subsection{Numerical analysis - Frequency Domain}

\begin{figure}[htbp] 
   \centering
   \includegraphics[width=0.4\columnwidth]{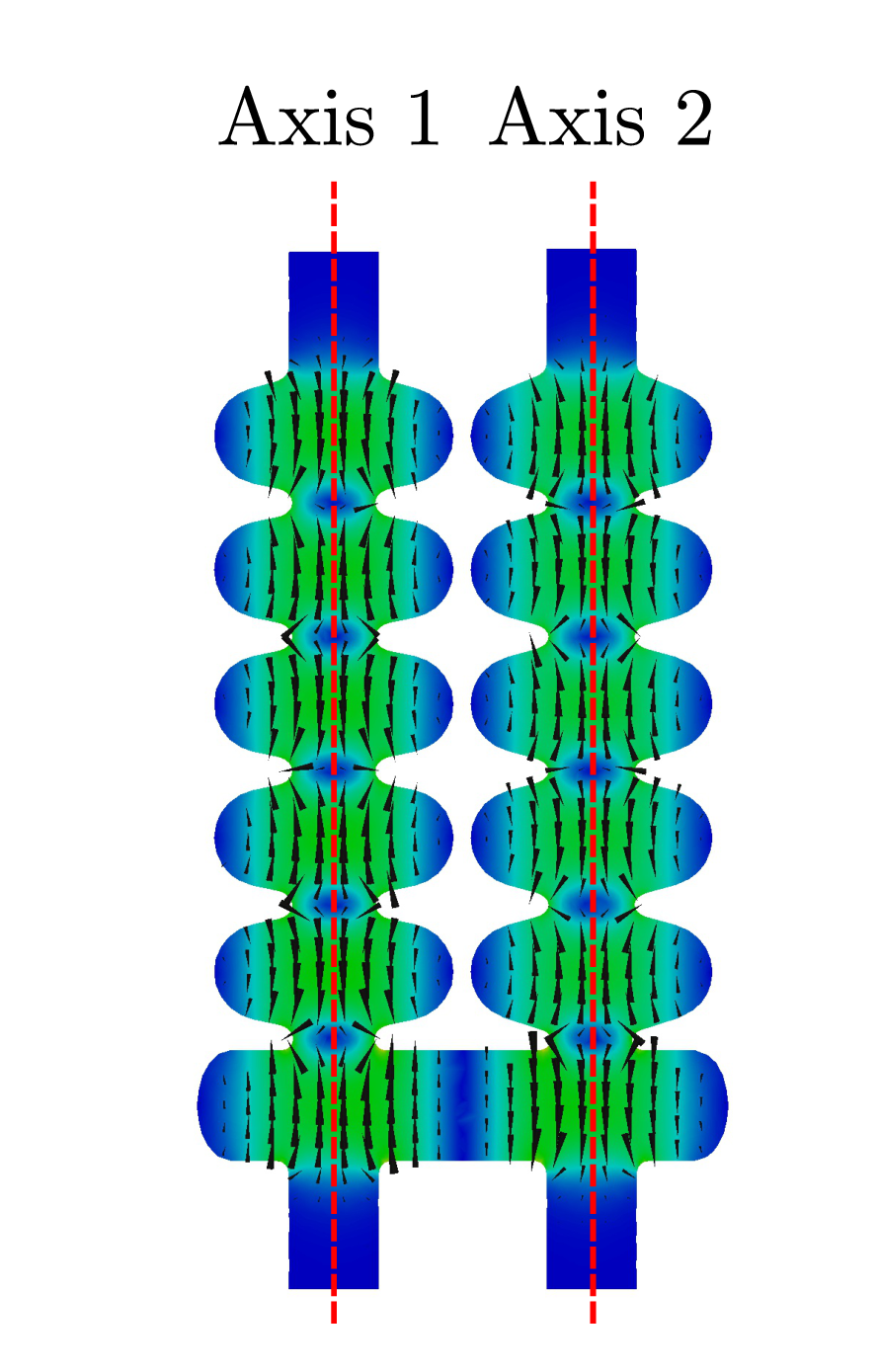} 
      \includegraphics[width=0.49\columnwidth]{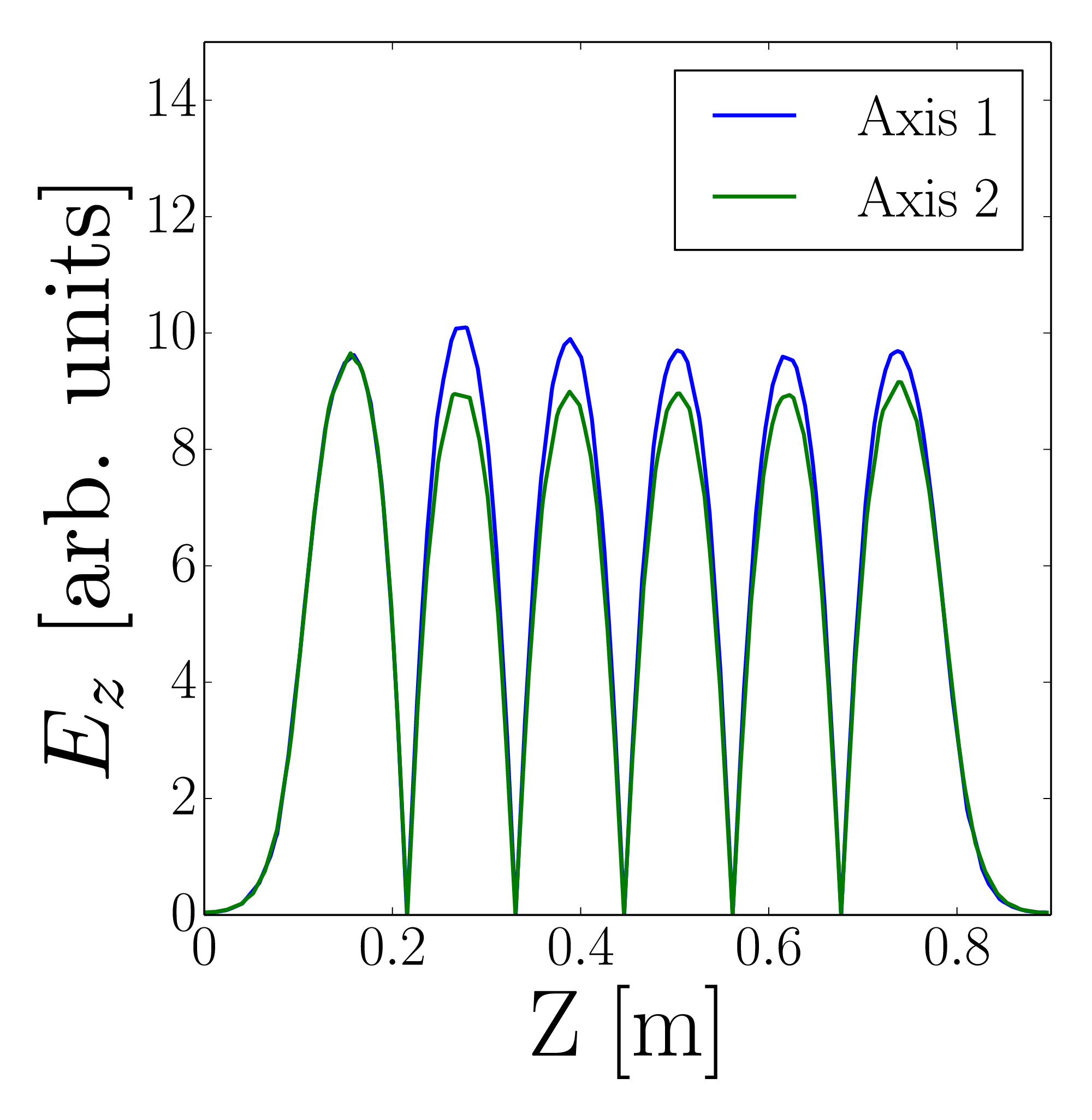} 

   \caption{A contour plot of the electric field distribution for the operating mode (left) and the electric field along each axis (right).}
   \label{fig:piMode}
\end{figure}

As we mentioned in the introduction to this chapter both types of analysis i.e. frequency and time domain were performed (for verification and illustration purpose) to study these structures. FIG. \ref{fig:piMode} shows the desired operating mode (at 1.3~GHz) transverse structure together with the Ez field flatness along the longitudinal axis z (FIG. \ref{fig:piMode}). The field flatness was calculated ignoring the coupling cell and for both axes is better than $98\%$ which is acceptable in most practical cases and can be further improved if required. It should be noted that the design for this paper provides a proof of principle and is by no means a fully optimised structure.

High efficiency SRF ERLs require very high quality factor of the fundamental mode. Thus, there will be a very strong sensitivity to vibration and pressure  known as "micro-phonics". To the operating mode the two cavities look like a single eleven-cell cavities rather than two coupled cavities, as the coupling cell is also resonant. This is no different than a normal nine cell cavity where the microphonics can shift the cell frequencies independently. Tuning becomes more difficult as the cavity gets longer as the modal frequencies in the passband come closer together, which is why the design shown here is limited to five cells on each side and strong coupling is used.

 Due to the different cell designs the amplitude of the electric field in each axis is slightly different (FIG. 6b) which can also be beneficial (i.e. possibility to vary the field in accelerating and decelerating cells) for energy recovery after beam interaction at the IP. We have calculated and analysed the first 100 eigen-modes of the system. For each mode, the complex longitudinal voltage $V_{\parallel,n}(r)$ was extracted in order to calculate the longitudinal $R/Q$ and transverse $R/Q$ for each axis.The longitudinal voltage was extracted using

\begin{equation}
V_{\parallel,n}(r) = \int^\infty_{-\infty} E_{z,n} (r,z) e^{i\omega_nz/c}\mathrm{d}z
\end{equation}

\noindent while the longitudinal $R/Q$ was calculated using

\begin{equation}
R/Q_n = \frac{\left|V_{\parallel,n}(0)\right|^2}{\omega_n U_n}
\end{equation}

Considering the fact that the voltage for a dipole mode scales linearly with $x$, the transverse voltage $V_{\perp,n}$ can be written using the longitudinal voltage as

\begin{equation}
V_{\perp,n} = i\frac{c}{\omega_n r}\left[V_{\parallel,n}(r)-V_{\parallel,n}(0)\right]
\end{equation}

In a similar way to the longitudinal $R/Q$, the transverse $R/Q_{\perp,n}$ can be defined as 

\begin{equation}
R/Q_{\perp,n} = \frac{\left|V_{\perp,n}(r)\right|^2}{\omega_n U_n}
\end{equation}

\begin{figure}[htbp] 
   \centering
   \includegraphics[width=1\columnwidth]{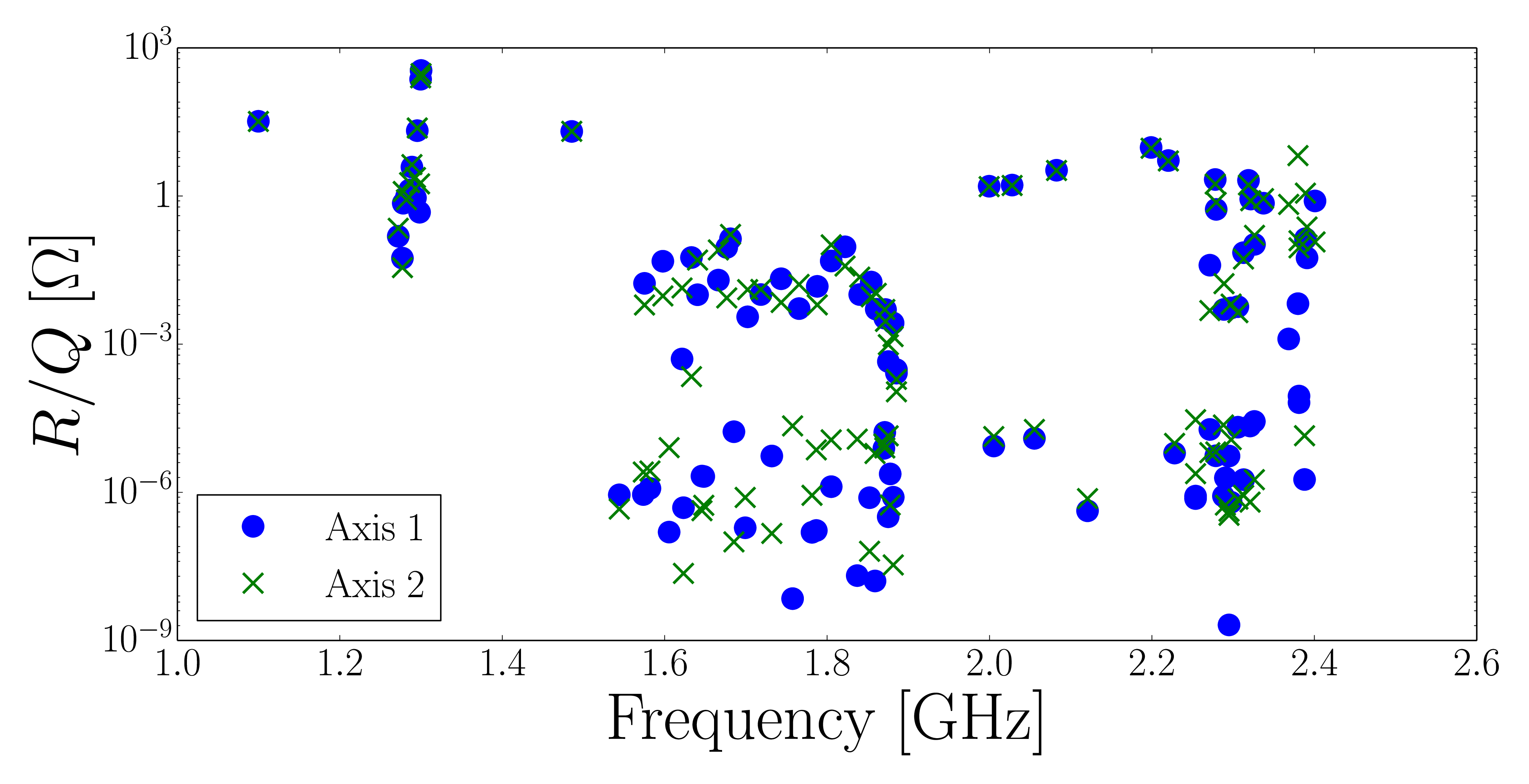} 
      \includegraphics[width=1\columnwidth]{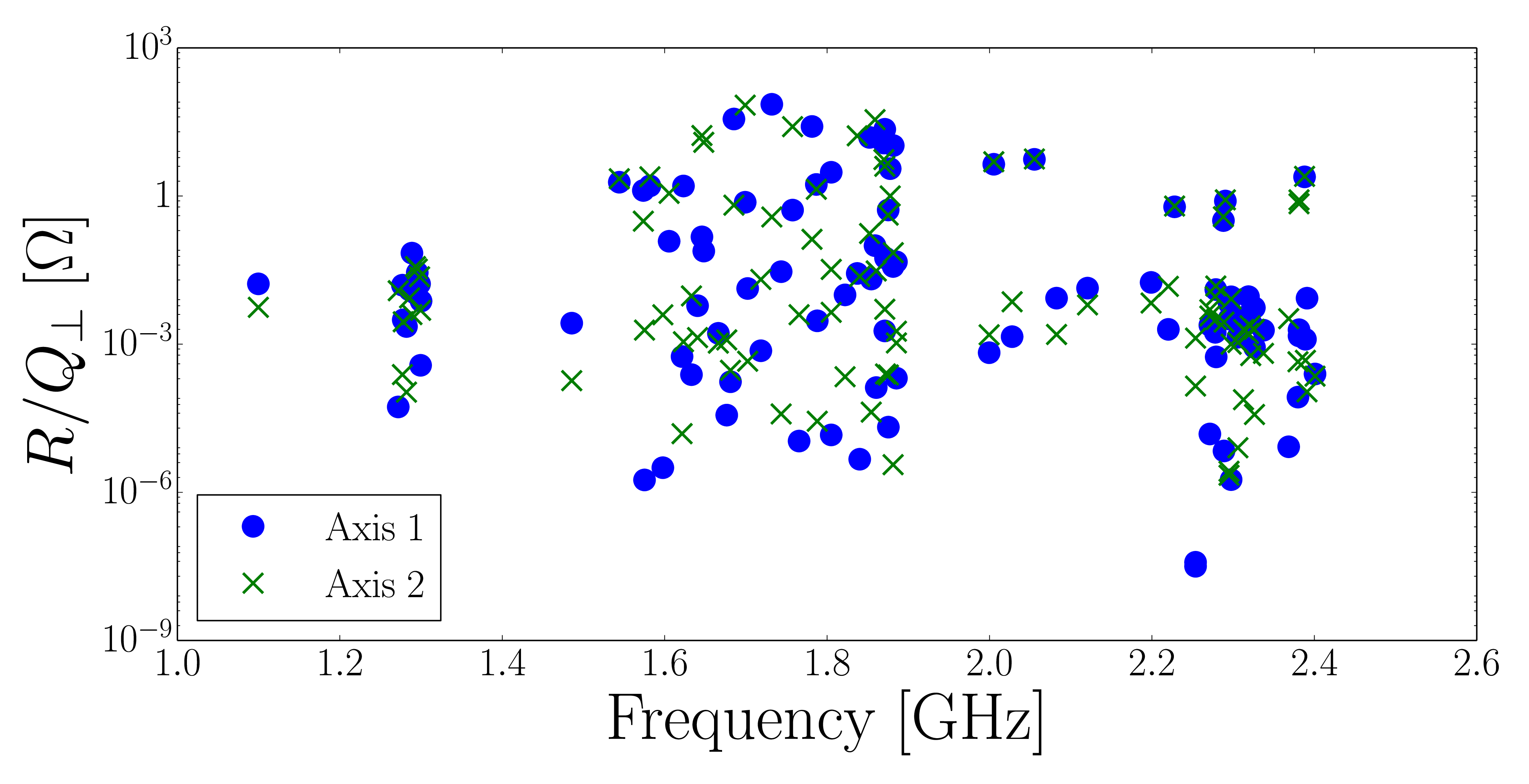} 

   \caption{The longitudinal and transverse $R/Q$ evaluated along both axes.}
   \label{fig:RQ}
\end{figure}

 The transverse $R/Q$ was evaluated at 1 mm offsets in the horizontal (x) and vertical (y) directions from the centre of each axis. FIG. \ref{fig:RQ} shows that the $R/Q$'s are not equal for each axis as desired. A cluster of modes is visible between 1 and 1.4 GHz. The majority of these modes lie inside the fundamental pass-band (around 1.3~GHz) and include the operating mode. On either side of this pass-band two modes have equal $R/Q$ (at $\sim$ 1.1~GHz and $\sim$ 1.5~GHz) are present. However these modes are confined/localised to the coupling cell and damping is thought to be possible. This will be illustrated and discussed below. The next cluster of modes has high transverse $R/Q_\perp$ suggesting these are dipole modes. These dipole modes having the highest $R/Q_\perp$ are of most concern for preventing the development of BBU.

\begin{table}[htbp]
   \centering
   \begin{tabular}{lcccccc}
   \hline
   \hline
    & Axis 1 & Axis 2 & Axis 1 & Axis 2 & Axis 1 & Axis 2 \\
   Frequency &\multicolumn{2}{ c }{$R/Q$}  &\multicolumn{2}{ c }{$R/Q_{\perp,x}$} &\multicolumn{2}{ c }{$R/Q_{\perp,y}$}\\
   GHz &\multicolumn{2}{ c }{$\Omega$} &\multicolumn{2}{ c }{$\Omega$} &\multicolumn{2}{ c }{$\Omega$} \\
   \hline
   \multicolumn{7}{ c }{Highest $R/Q$} \\
   \hline
1.3 & 348.71 & 301.51 & 0.0675 & 0.0365 & 0.0074 & 0.0 \\
1.29943 & 231.71 & 247.59 & 0.0014 & 0.0059 & 0.0003 & 0.0048 \\
1.09966 & 32.622 & 32.367 & 9.5769 & 9.0660 & 0.0166 & 0.0055 \\
1.29532 & 21.075 & 23.878 & 0.0014 & 0.0267 & 0.0281 & 0.0333 \\
1.48554 & 20.337 & 20.429 & 12.094 & 12.360 & 0.0026 & 0.0001 \\
\hline
   \multicolumn{7}{ c }{Highest $R/Q_{\perp,x}$} \\
\hline
1.70216 & 0.0035 & 0.0127 & 65.207 & 0.8680 & 0.0134 & 0.0004 \\
1.74343 & 0.0211 & 0.0069 & 61.997 & 0.4792 & 0.0294 & 3.8679 \\
1.87193 & 0.0050 & 0.0029 & 35.500 & 0.0810 & 0.0555 & 0.0002 \\
1.85436 & 0.0181 & 0.0091 & 17.329 & 0.3260 & 0.0208 & 4.2119 \\
1.48554 & 20.337 & 20.429 & 12.094 & 12.360 & 0.0026 & 0.0001 \\
\hline
   \multicolumn{7}{ c }{Highest $R/Q_{\perp,y}$} \\
\hline
1.73192 & 5.4419 & 1.4736 & 0.0005 & 0.0001 & 72.089 & 0.3764 \\
1.68526 & 1.6890 & 9.9178 & 0.0024 & 1.8274 & 36.312 & 0.6537 \\
1.78142 & 1.5499 & 8.7076 & 0.0039 & 0.0070 & 25.636 & 0.1329 \\
1.87103 & 1.6368 & 7.9211 & 4.8525 & 0.0037 & 22.491 & 4.0005 \\
1.8523 & 7.7902 & 6.4131 & 0.0033 & 0.0001 & 15.388 & 0.1740 \\
\hline
\hline
   \end{tabular}
   \caption{The modes with the top five highest $R/Q$, $R/Q_{\perp,x}$ and $R/Q_{\perp,y}$.}
   \label{tab:RQvalues}
\end{table}

\begin{figure}[htbp] 
   \centering
   \includegraphics[width=1\columnwidth]{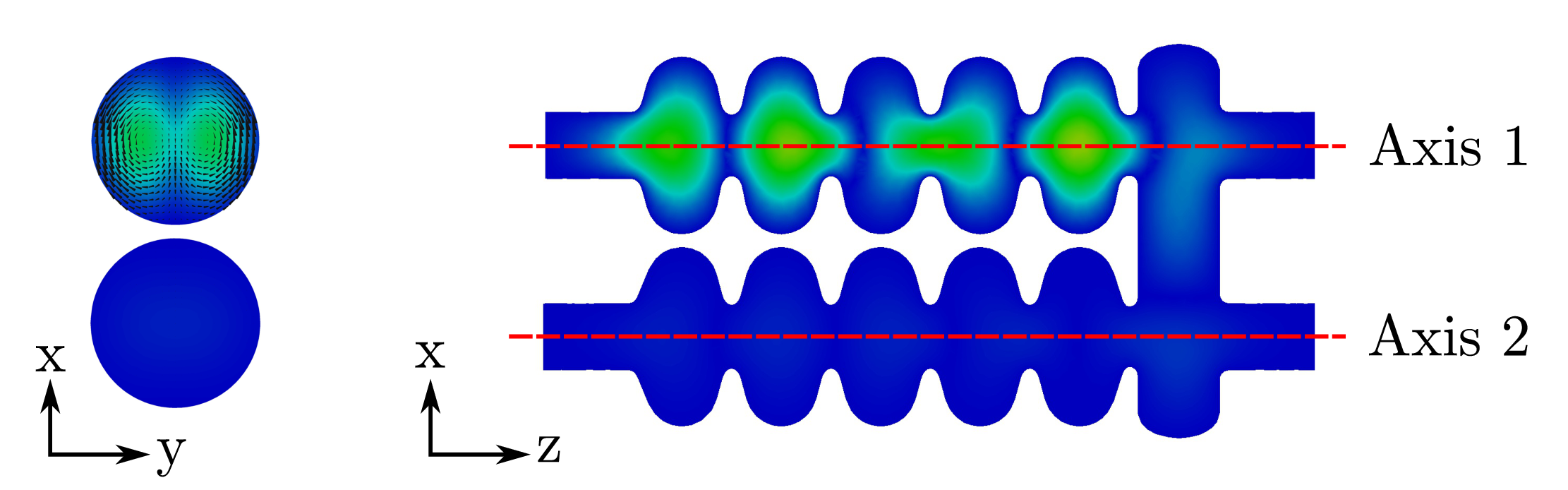} 
   \caption{A transverse and longitudinal slice showing the electric field contour plots for a mode at 1.73 GHz. The transverse slice also shows the magnetic field as indicated by the cones.}
   \label{fig:worst}
\end{figure}

Looking at the worst scenario we note that the transverse $R/Q$ of one of the mode has an $R/Q$ of 72~$\Omega$. In conventional system this mode would have the same $R/Q$ along the second axis leading to BBU development. However in this case the $R/Q$ along the second axis is only 0.38~$\Omega$. This confirms that  due to the asymmetric design of the system it works as expected, namely confining the HOM to just one of the axis. FIG. \ref{fig:worst} shows a transverse and longitudinal slice of this mode. 

\begin{figure}[htbp] 
   \centering
   \includegraphics[width=0.5\columnwidth]{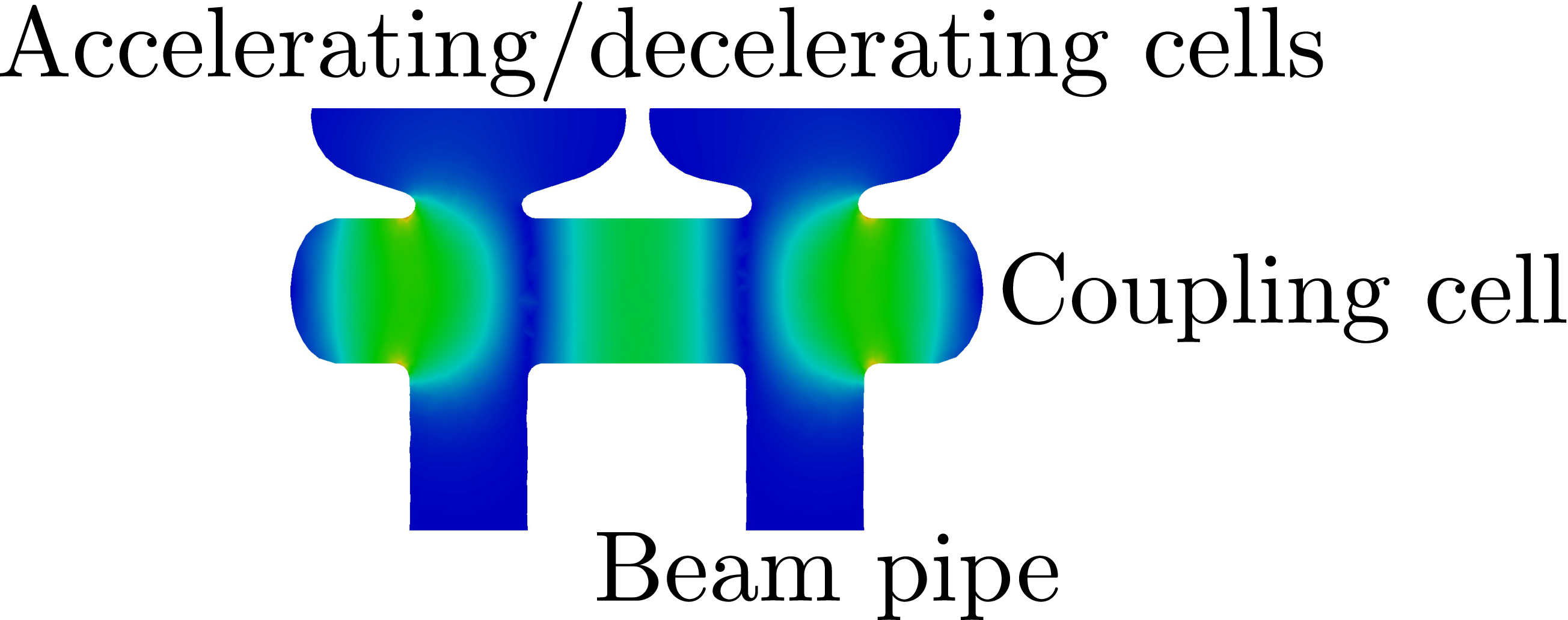} 
   \caption{A mode confined to the coupling cell which is symmetric.}
   \label{fig:CCmode}
\end{figure}

As mentioned above and shown in FIG. \ref{fig:RQ} there are modes (due to symmetric shape of the coupling cell) which have the same $R/Q$ along each axis. These modes are confined to this cell (FIG. \ref{fig:CCmode})  resulting in the $R/Q$ being equal along each axis but also meaning that such modes can be easily dealt with by applying an appropriate coupler or an absorber. An example of such a mode (located at frequency 1.4855 GHz)  is shown in FIG. \ref{fig:CCmode} and it is clear that locating a coupler at the middle of the coupling cell (for instance) will not affect the beam transportation and will effectively damp this mode.

\subsection{Numerical analysis - Time Domain}

One of the limiting factors of frequency domain calculations is the use of closed boundary conditions. An alternative is to perform a time domain calculation in which the beam pipe boundaries allow waves above the cutoff frequency of the beam pipe to propagate. The fields generated by a charge on a trailing particle can also be easily calculated and compared/verified with the data observed from frequency domain analysis. 

To study wake-field excitation let us consider a Gaussian bunch with charge density
\begin{equation}
\rho(s) = \frac{1}{\sqrt{2\pi}\sigma}e^{-\frac{(s-s_0)^2}{2\sigma^2}}
\end{equation}

\noindent where $\sigma$ is the bunch length (we assumed that $\sigma = 1~cm$), $s$ is the distance along the path. The bunch is then launched along one axis of the system with a vertical offset assumed to be 1~mm. By this way the full 3D model is simulated meaning both monopole and dipole like modes are excited. The simulations were repeated with the bunch traveling along the second axis in the opposite direction. The longitudinal wakefield was calculated using Weiland's indirect scheme \cite{Weiland:1980qc} and the wake-field potential was Fourier transformed/analysed to obtain the complex impedance spectrum using

\begin{equation}
Z_{\parallel} (\omega) = \frac{1}{c\rho(\omega)}\int^\infty_{-\infty} W_{\parallel} e^{-i\omega s/c}\mathrm{d}s
\end{equation}

\begin{figure}[htbp] 
   \centering
   \includegraphics[width=1\columnwidth]{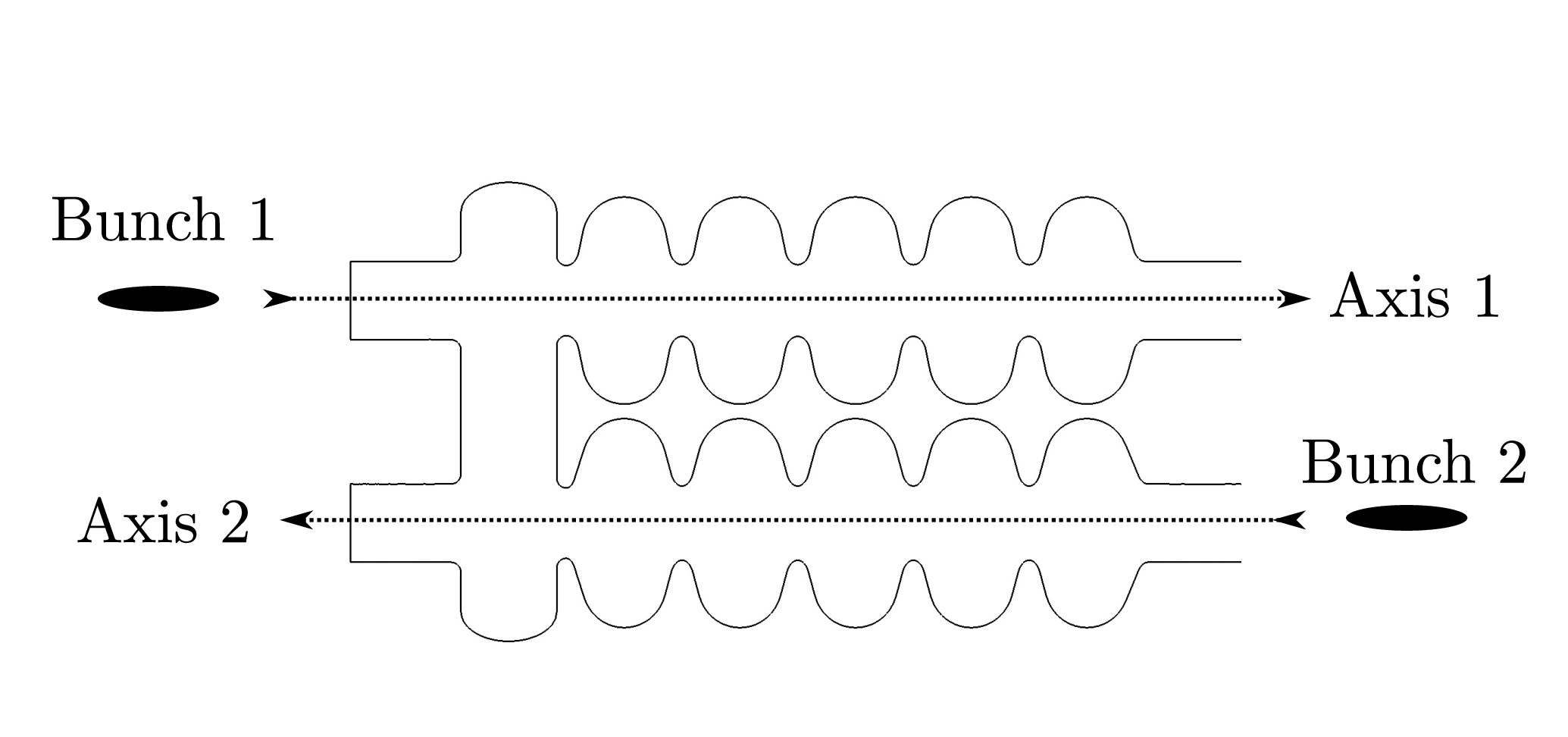} 
   \caption{A schematic of the simulations setup.}
   \label{fig:TDsim}
\end{figure}

\begin{figure}[htbp] 
   \centering
   \includegraphics[width=1\columnwidth]{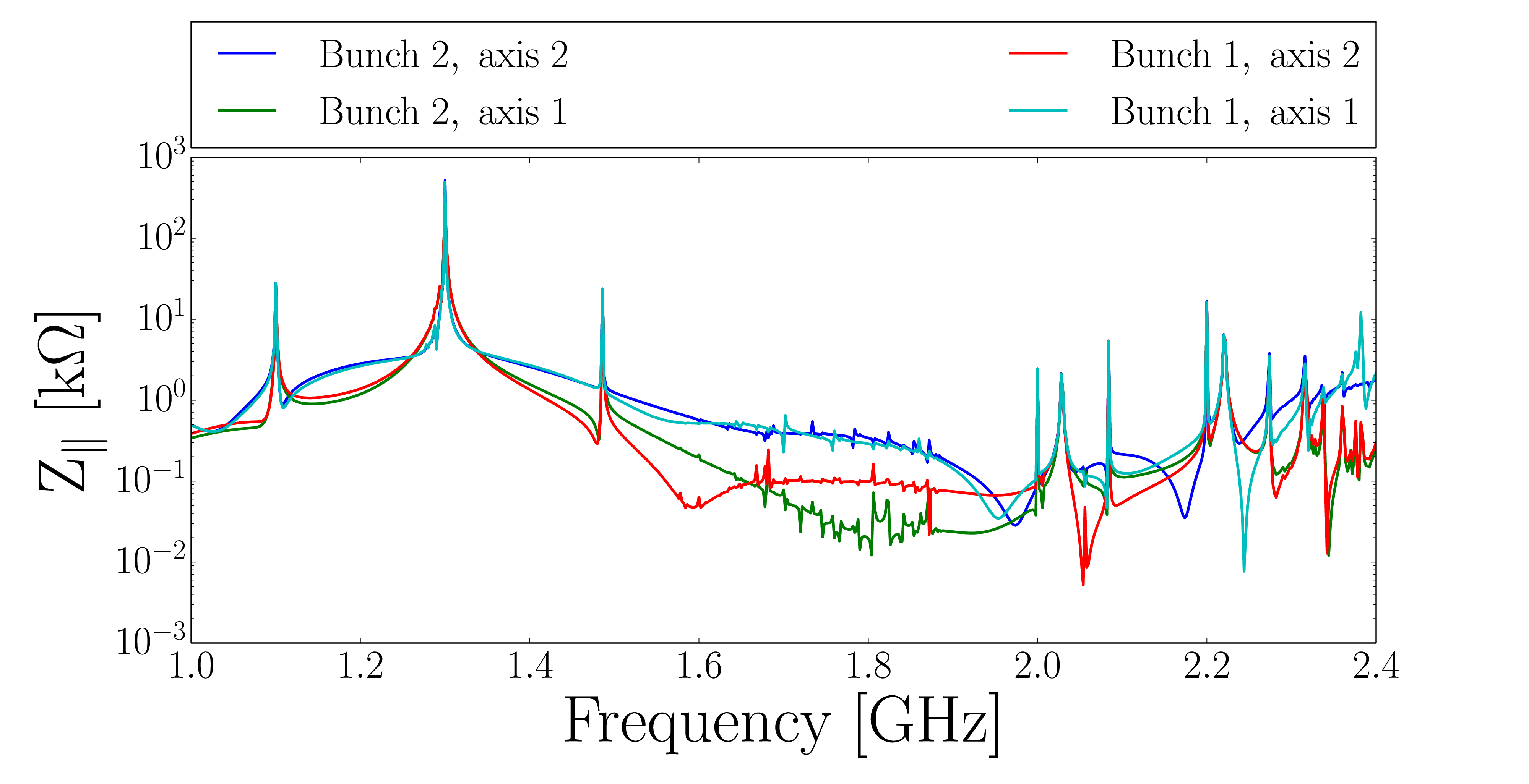} 
      \includegraphics[width=1\columnwidth]{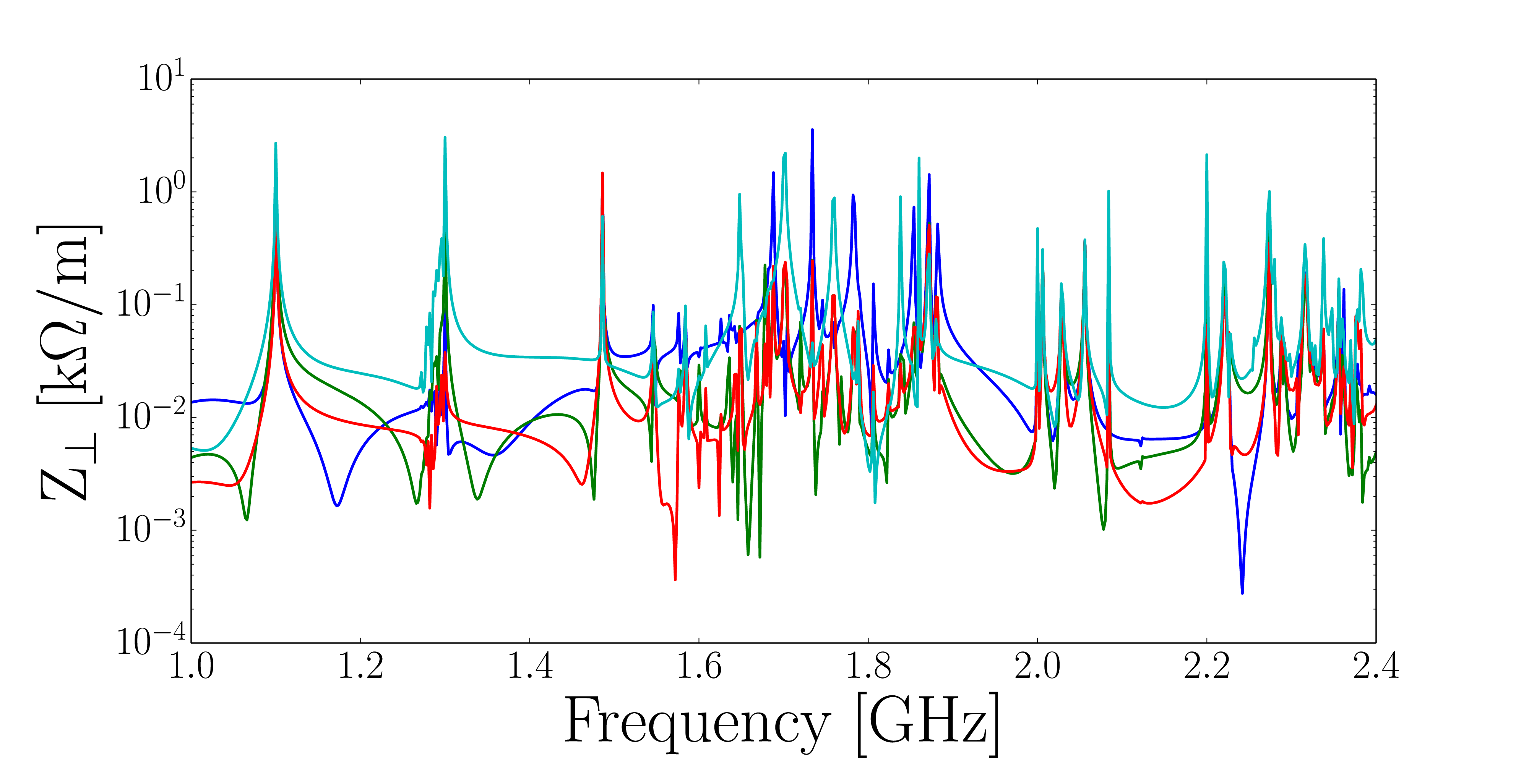} 

   \caption{The longitudinal impedance (top) and transverse impedance (bottom) spectrum.}
   \label{fig:Z}
\end{figure}

The Panofski-Wenzel theorem \cite{p-w} was then applied in order to calculate the transverse impedance spectrum

\begin{equation}
\frac{\omega}{c} Z_{\perp} (\omega) = \nabla_{\perp} Z_\parallel (\omega)
\end{equation}

\noindent where $\nabla_\perp = \frac{\partial}{\partial x}\vec{x} + \frac{\partial}{\partial y}\vec{y}$. The schematics of simulations discussed  is  shown in FIG. \ref{fig:TDsim}. The two simulations are similar but  in the first simulation, bunch 1 is launched along axis 1 in the positive Z direction and the wake is calculated along axis 1 and 2 with a frequency resolution of 2~MHz. The simulation is then repeated (model 2) for the opposite case (with the same resolution) where bunch 2 is launched along axis 2 in the negative Z direction.  In both models a time-step of 2~ps  was used to observe the convergence of the results. The spectra of longitudinal $Z_{\parallel ij}$ and transverse $Z_{\perp ij}$ impedance observed from these simulations are shown in FIG. \ref{fig:Z}. The subscripts $ij$ indicate in both cases the number of axis and the bunch as discussed. Let us note a good agreement between FIG \ref{fig:RQ} and \ref{fig:Z}, indicating a good convergence of the results observed from the time and the frequency domain approaches. One can easily identify different families of modes and see a good correlation between the amplitude of impedances and $R/Q$ parameters of the modes. It is also interesting to note that the spectra of the impedances of the wakefields associated with HOMs (frequencies above 1.5~GHz) measured on different axes are different (light and dark blue lines can be easily distinguished in the figure).

Indeed it can be seen from FIG. \ref{fig:Z} that the operating mode's impedances (1.3 GHz) are roughly equal. A number of other modes can been seen with almost equal coupling to the beam including a coupling cell modes (equidistantly located on either side of the operating mode) and a band starting at 2 GHz. The last group of the modes is also associated with the symmetric coupling cell and can be easily damped if necessary. A band of cavity modes which are of our concern is located between 1.6 and 1.9~GHz. Their impedances spectra vary as desired and thus they may not be considered as a threat to the system stability. Also these modes are dipole like modes and thus their contribution is better shown in the transverse impedance spectrum in FIG \ref{fig:Z}.

The transverse impedance spectrum still shows the operating mode but to a much smaller degree. The effect of the cavity modes between 1.6-1.9 GHz is now much more visible. As desired, the frequency and coupling of these modes are different depending on the axis. There is now a lower order mode at 1.1 GHz which is the monopole mode of the coupling cell. It has a large impedance, is symmetric and as the beampipes are not at the centre, it will have a transverse component.

\section{Conclusions}\label{sec:conclusions}

\begin{figure}[htbp] 
   \centering
   \includegraphics[width=1\columnwidth]{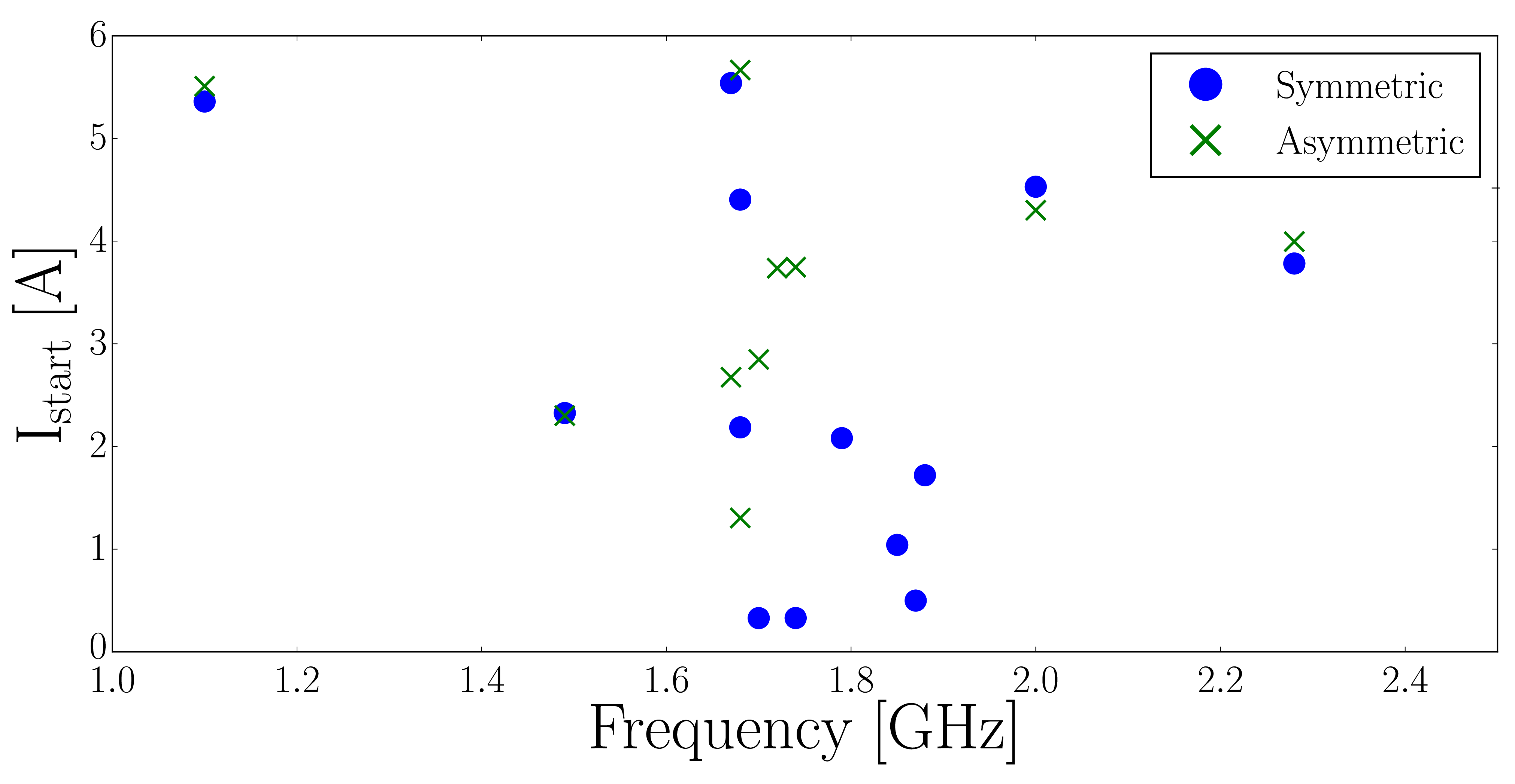} 
   \caption{Comparison of the start currents for symmetric and asymmetric cavities is shown using $T_g=2\pi n$}
   \label{fig:istart1}
\end{figure}

In order to evaluate the benefits of such an asymmetric cavity we calculate the start current for a sample machine with an injection energy of 5~MeV and a cavity voltage of 5~MeV. We will assume an $\mathbf{R}_{12}=1~$m. It is necessary to calculate both the BBU start current (calculated in section \ref{sec:numericalModelling}) as well as current at which the wakefield drives the bunch into a wall due to the electrical centre shift of the dipole mode (as calculated in section \ref{sec:rlc}).
The BBU start current can be calculated for each mode in the cavity, and the lowest value is taken as the start current for the machine. Let us first calculate the worst case ($\phi=\pi/2$) and  $\omega T_g=n\pi+\pi/2$ where $n$ is an integer. In FIG \ref{fig:istart1} the comparison of the start currents for symmetric and asymmetric cavities is shown using $\omega T_g=2\pi n$. As we are only concerned with the modes with the lowest start currents we only show modes with start currents below 6~Amps. As can be seen the start current increases by a factor of 4.

\begin{figure}[htbp] 
   \centering
   \includegraphics[width=1\columnwidth]{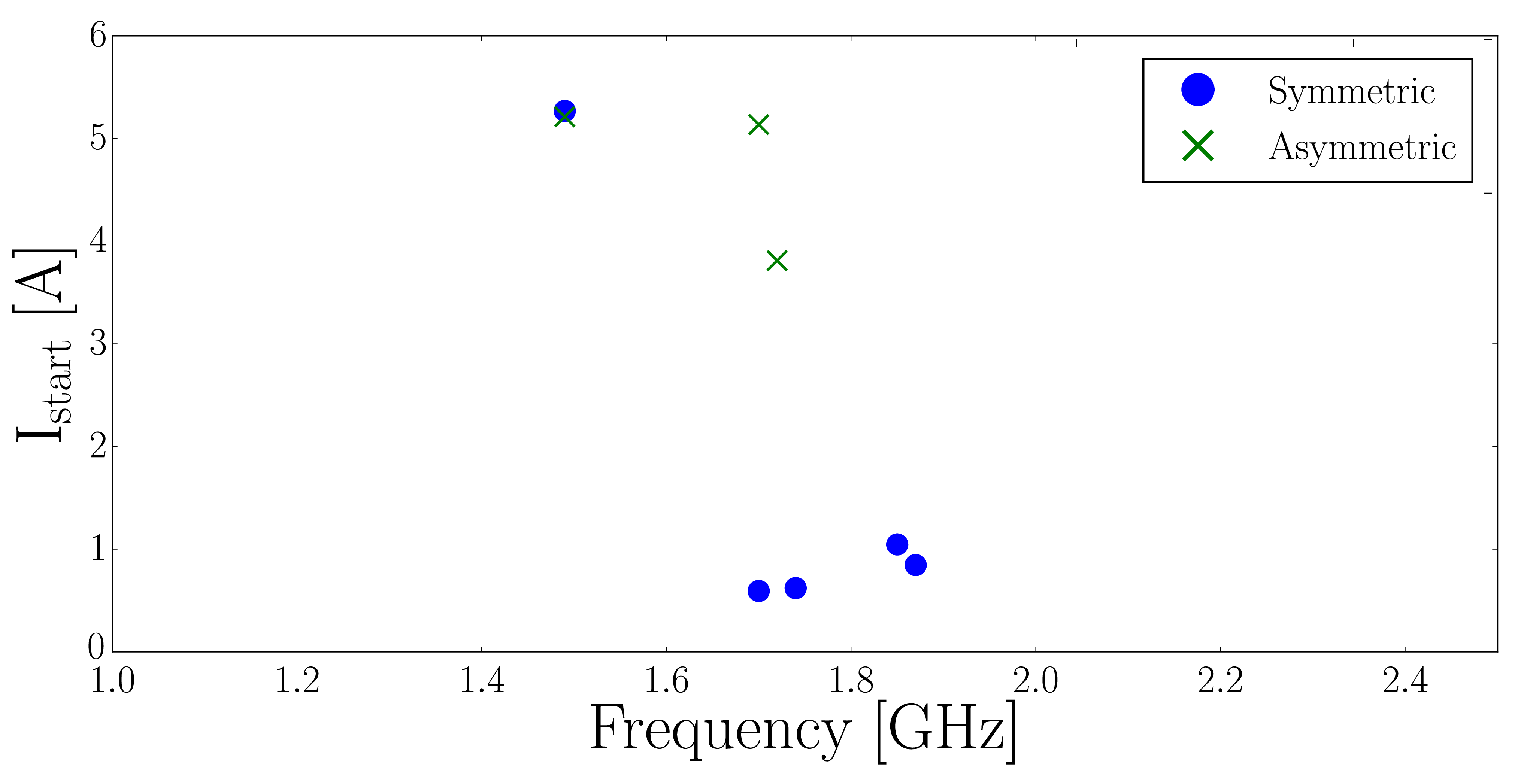} 
   \caption{Comparison of the start currents for symmetric and asymmetric cavities is shown using $T_g=7.69~ns$.}
   \label{fig:istart2}
\end{figure}

As a next step, one can look at the case where $T_g$ is fixed. In order to maximize the energy recovery process, the delay time must be an integer multiple of the RF period at the operating frequency. Here we use the 10th sub-harmonic of 1.3~GHz giving a delay of 7.69 ns which is a path of just over 2 m. In FIG \ref{fig:istart2} the comparison of the start currents for symmetric and asymmetric cavities is shown for such a delay $T_g$=7.69 ns. The start current observed for the asymmetric machine of 3.8~A is dominated by the symmetric dipole mode at 1.72~GHz (eigenmode of the symmetric coupling cell discussed in section 4), limiting the benefits of the asymmetric cavity. However, this mode has a smaller $R/Q$ than the operating dipole modes and also this mode can be effectively dealt with by either providing strong HOM damping or by altering the cells such that the sum wake has a phase which is an integer multiple of $\pi$. These methods are viable for the asymmetric system as only a single mode needs to be considered. As can be seen, the start current for all other modes is significantly higher for an asymmetric machine (3.8~A) compared to a symmetric machine (593~mA) by a factor of 5. One may argue that due to the asymmetry of the cavity the dipole modes electrical centre doesn't align with the design orbit leading to the possibility of the induced wakefield to become large enough to drive the bunch into a wall even without a feedback path. However, using the equations derived in this paper one finds that the current at which the wake kicks the beam to a significant offset to dump the beam, i.e. 10~cm, is over 1000~A, and hence is not a limit to the machines maximum operating current.

If the coupling cell were to be redesigned it could be possible for the start current to tend to infinity. One could design a coupling cell which would make use of magnetic coupling. However, a magnetic coupling cell would require a weld at a high current region and such approach was discounted, at present, for this design. For the design presented here, the coupling cell is also symmetric. An antisymmetric coupling cell could also be designed to increase the start current threshold further.

If the transit time is ignored, the limiting factor is driven by symmetric coupler modes however, in the case when the transit time is included, the length of the arc connecting the acceleration and deceleration arm can be chosen such the beam traverses the deceleration arm in anti-phase with respect to a symmetric mode, thus avoiding BBU. The limiting factor then arrises from the fact the $N_{SP}$ is non-zero and thus, the BBU limit will be raised compared to a symmetric structure but not infinite. The reason $N_{SP}$ is non-zero is because both cavities sections are coupled, eigenmodes will be of the full system and not confined to one arm. Modes that have high $N_{SP}$ are probably caused by two modes very close in frequency space such that there is some coupling.

To sum up: in this paper we introduced a new concept of asymmetric system of accelerating and decelerating cavities coupled by a resonant coupler for ERL. We considered both numerical and analytical approaches and demonstrated the operational principles of new system. We have compared it with conventional symmetric design and discussed the advantages of the asymmetric layout. Indeed it was demonstrated that asymmetry allows transporting through the ERL an electron bunch having significantly increased (at least by factor of 5) current without limiting BBU. This potentially is very attractive for applications in which it is important to increase THz and x-ray photon yield as well as bunch energy recovery.


\bibliography{apssamp}

\appendix
\section{Effective transverse potential due to multiple bunch interaction} \label{app:multilaunch}

From the previous discussion it is clear that bunch deflection and time delay are affected by the effective transverse potential which we discussed for a single bunch in the sections above. Therefore, to calculate $\Delta x$ of bunch N+1 after N bunches passed through the system or the bunch delay at the decelerating section, the transverse kick due to field accumulation inside the cells needs to be estimated. We will note that $E_z^n (r,\phi)$  are the functions which define TM$_{on}-$ like and TE$_{on}-$  like modes' transverse structures. To derive the  transverse deflection, expression (\ref{eq:deviation}) will be used and we will consider the worst scenario i.e. maximum transverse kick. We will also consider HOMs of a single un-coupled cavity with the strongest contribution to the  $\theta$. This is arisen from the fact that the asymmetric structure is made of two cavities which are weakly coupled, away from the resonant frequency of the coupler. For the purpose of discussion, we use only the first few dipoles HOMs which have the highest impedance and hence dominates the long range wake allowing us to ignore the rest of the high frequency HOMs.  We note that in the model described, after energy recuperation stage, the bunch is dumped into a collector. Due to this the system has the following specific times: repetition rate $T_{rep}$ , time decay $T_{dec}$ of HOM (we assume that Q-factors of respective HOMs are the same in both cavities), and the bunch transit time, $t_0$ between the accelerating and decelerating arms. Let us now define $T_{dec} > T_{rep}$ and $T_{dec} >t_0$ while taking into account that all these times are much larger as compared with bunch transient time T through the structure. We note that a field amplitude seen by the following (second) bunch will decrease in time (dissipation losses due to finite value of quality factor Q for HOMs): 

\begin{equation}
\left|V_2\right| = \left|V_1 e^{-T_{rep}/T_{dec}}\right|
\end{equation}

\noindent $V_{1,2}$ are the voltages generated (3) by the first (driver) and seen by the witness bunch respectively and $1/T_{dec} =\omega/2Q=\alpha$ is the decay factor. Let us now consider a multi-bunch excitation of the wake-field taking into account a linear accumulation of the fields, i.e. yield from each bunch is the same, while taking into account the phase and the Q-factor of the cavity eigenmodes and ignoring non-linear effects which may exist. To start we use the monochromaticity of the field and modify the expression (6) to evaluate the multi-bunch effect $V_{\perp, multi}$:

\begin{equation}\label{eq:vperpmulti}
-ic\int^T_0 e^{i\omega\tau} \mathrm{d}\tau\int^L_0 \nabla_\perp \left[E_z(x, z) \sum\limits^N_{n=1} e^{(i\omega - \alpha)nT_{rep}}\right] \mathrm{d}z
\end{equation}

\noindent Here we use cylindrical coordinates to describe fields only, $E_z$ is generated by the bunches and seen by the witness bunch. In (\ref{eq:vperpmulti}) the introduction of `discrete' time for EM field accumulation i.e. proportional to $T_{rep}$ is possible due to very short transient time through the structure $T_{rep} \gg T$. To analyse the field accumulation we can introduce a decay $\alpha$ and detuning $\delta \omega_n$ parameters  $\alpha=1/T_{dec}$, $\omega=2\pi \tilde{n} /T_{rep} +\delta \omega_n$  where $\tilde{n}$ is an integer number indicating a number of full oscillation which EM field does at the operating frequency before next bunch enters the cavity. Clearly $\tilde{n}$ is not connected to n (number of bunches) and can be either arbitrary large or small (large or  $\tilde{n} \rightarrow \infty$ is the case of single bunch approximation). Taking only a real part of the sum (\ref{eq:vperpmulti}) the  expression can be rewritten as:  $\sum\limits^N_{n=1} e^{-nT_{rep}/T_{dec}}\sin (n\delta\omega_n T_{rep})$ leading to the expression 

\begin{gather}
\begin{aligned}\label{eq:vperpmultimax}
(V_{\perp,multi})^{max} &\le \\ 
-F^{\infty}& \int^T_0 e^{i\omega\tau}\mathrm{d}\tau\int^L_0 \nabla_\perp E_z (x, z) \mathrm{d}z
\end{aligned}
\end{gather}

\noindent where \cite{Burt:2007zzd}

\begin{equation}
F^\infty =\frac{\sin (\delta\omega_n  T_{rep})}{2\cosh (T_{rep}/T_{dec})-\cos (\delta\omega_n  T_{rep})]}
\end{equation}

We note that the expression (\ref{eq:vperpmultimax}) can be used to find the deflection and time delay using the approach developed for a single bunch. However, it requires detailed knowledge of $E_z$ which can be obtained only with very accurate numerical analysis. To make reasonable estimations the following steps can be taken. One notes that if the driving bunches with transverse offset $r$ from the mode electrical centre is acting upon trailing (by distance $z$ or time interval $t=z/c$ (not $\tau$)) charges, it will provide a transverse kick via the excitation of the transverse dipole mode. Assuming a multipole representation as discussed above (\ref{eq:deltar_pillbox}) one can define d$z=$cd$t$ and $\nabla_\perp [E_z (r,\phi,z)]$ in (\ref{eq:vperpmultimax}) as:

\begin{equation}
V_\perp (r, t) = 2qr\sum\limits^M_{m=1} K_m \sin (\omega_m t) e^{-\frac{\omega_mt}{2Q_m}}
\end{equation}

\noindent where $\omega_m,Q_m$ are the frequency and the Q factor of multipoles with index $m$ (monopole modes are not considered as they will not contribute to the kick). The amplitudes $K_m$ of multipoles' potential is also known as ``kick factors'' and are  given by $K_m=c/4 [R/Q]_j^m$. Applying (\ref{eq:RQintro}) and taking into account only the first dipole mode m=1 (i.e. ignoring high-order multipoles) $K_m$ can be presented as

\begin{equation}\label{eq:kick} 
K_1=\frac{c\left|V_1^j(x)\right|^2}{4\omega_1r^2U_1}
\end{equation}

\noindent where $r$ indicates the transverse deviation from the designed trajectories in accelerating (superscript j=1) and decelerating (superscript j=2) cavities, $U_m$ is the total energy of the mode, and $\omega_m$ is the mode eigen-frequency. To study the damping requirements, which ensure the beam does not exceed the specified deflection, we undertake a single-mode analysis and hence we will drop the subscript m or, in case of (\ref{eq:kick}), subscript 1. Following from (\ref{eq:rtplust0sim}-\ref{eq:deltar_0}), and applying expressions (\ref{eq:vperpmulti}-\ref{eq:kick}) observed to (\ref{eq:deltar_pillbox}-\ref{eq:r12cond1}), we estimate radial offset between the beam trajectory and designed pass

\begin{equation}
V_{\perp,max} = 2Ir K_mF^\infty / \omega
\end{equation}

\noindent Taking into account (\ref{eq:deltar_0}) with $\Delta r = D_2$ one finds that the expression for the maximum current:

\begin{equation}
I_{max} = (D_2 W\omega)/(2x K_mF^\infty)
\end{equation}

\end{document}